\documentclass{aa} 
\usepackage{times,psfig,epsfig,amssymb,amsmath}

\newcommand{\chandra}{{\it Chandra} }

\newcommand{\sax}{{\it BeppoSAX} }
\newcommand{\rosat}{{\it ROSAT} }
\newcommand{\xmm}{{\it XMM-Newton} }

\title{Constraining the cosmological parameters with the gas mass fraction 
in local and $z>0.7$ Galaxy Clusters}
\titlerunning{Cosmological constraints from the Cluster Gas Fraction}

\author{S. Ettori \inst{1} \and P. Tozzi\inst{2} \and P. Rosati \inst{1}}
\authorrunning{S. Ettori et al.}

\institute{
 ESO, Karl-Schwarzschild-Str. 2, D-85748 Garching, Germany
 \and INAF, Osservatorio Astronomico di Trieste, via Tiepolo 11, I-34131 Trieste, Italy
}

\offprints{S. Ettori}

\mail{settori@eso.org}

\date{Accepted on 14th Nov 2002}

\begin{document}

\abstract{
We present a study of the baryonic fraction in galaxy clusters 
aimed at constraining the cosmological parameters $\Omega_{\rm m}$,
$\Omega_{\Lambda}$ and the ratio between the pressure and density of
the ``dark'' energy, $w$.  We use results on the gravitating mass
profiles of a sample of nearby galaxy clusters observed with
the \sax X-ray satellite (Ettori, De Grandi \& Molendi, 2002) to set
constraints on the dynamical estimate of $\Omega_{\rm m}$.  We then
analyze \chandra observations of a sample of eight distant clusters 
with redshift in the range 0.72 and 1.27 
and evaluate the geometrical limits on the cosmological
parameters $\Omega_{\rm m}$, $\Omega_{\Lambda}$ and $w$ by requiring
that the gas fraction remains constant with respect to the look-back
time.  By combining these two independent probability distributions and
using {\it a priori} distributions on both $\Omega_{\rm b}$ and $H_0$
peaked around primordial nucleosynthesis and HST-Key Project results
respectively, we obtain
that, at 95.4 per cent level of confidence, (i) $w < -0.49$, (ii)
$\Omega_{\rm m} = 0.34^{+0.11}_{-0.05}$, $\Omega_{\Lambda} = 1.30^{+0.44}_{-1.09}$
for $w=-1$ (corresponding to the case for a
cosmological constant), and (iii) $\Omega_{\rm m} = 1-\Omega_{\Lambda}
= 0.33^{+0.07}_{-0.05}$ for a flat Universe.
These results are in excellent agreement with the cosmic concordance scenario
which combines constraints from the power spectrum of 
the Cosmic Microwave Background, the galaxy and cluster distribution, 
the evolution of the X-ray properties of galaxy clusters and 
the magnitude-redshift relation for distant type Ia supernovae.
By combining our results with the latter method we further constrain
$\Omega_{\Lambda} =0.94^{+0.28}_{-0.32}$ and $w < -0.89$
at the $2 \sigma$ level.
\keywords{galaxies: cluster: general -- galaxies: fundamental
parameters -- intergalactic medium -- X-ray: galaxies -- cosmology:
observations -- dark matter.}  }

\maketitle

\section{INTRODUCTION}

Several tests have been suggested to constrain the geometry
and the relative amounts of the matter and energy constituents of the 
Universe (see recent review in Peebles \& Ratra 2002 and references
therein). A method that is robust and complementary to the others is
obtained using the gas mass fraction, $f_{\rm gas} =
M_{\rm gas}/M_{\rm tot}$, as inferred from X-ray observations
of clusters of galaxies. In this work, we consider
two independent methods for our purpose: 
(i) we compare the relative amount of baryons with respect to the
total mass observed in galaxy clusters to the cosmic baryon fraction
to provide a direct constraint on $\Omega_{\rm m}$
(this method was originally adopted to show the crisis
of the standard cold dark matter scenario in an Einstein-de Sitter 
Universe from White et al. 1993), 
(ii) we limit the parameters that describe the geometry of the
universe assuming that the gas fraction is constant 
in time, as firstly suggested by Sasaki (1996).

The outline of our work is the following.
In Section~2, we describe the cosmological framework that allows
us to formulate the cosmological dependence of the cluster gas mass
fraction. In Section~3, we use a sample of nearby clusters to
constrain mainly the cosmic matter density. Eight galaxy clusters
with $z>0.7$ are then presented and analyzed in Section~4 and 
a further constraint on the geometry of the Universe is given under
the assumption of a constant gas fraction as function of redshift.
A description of the systematic uncertainties that affect our
estimates is discussed in Section~5. Finally, the combined
probability function and the overall cosmological constraints
(also considered in combination with results from SN type Ia
magnitude-redshift diagram) are described in Section~6 along with
prospective for future work. 

\begin{figure*}
\hbox{
\epsfig{figure=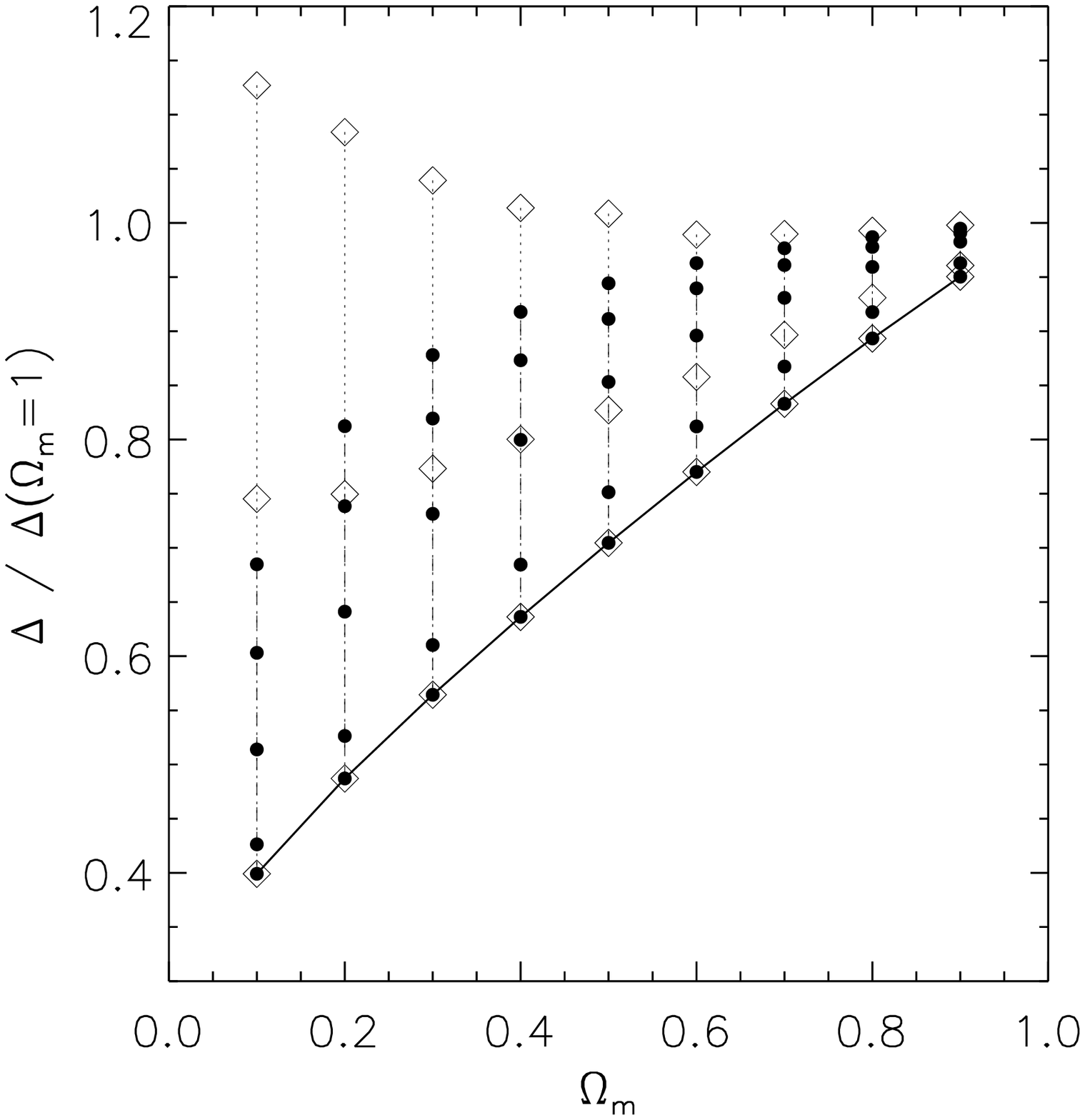,width=0.5\textwidth}
\epsfig{figure=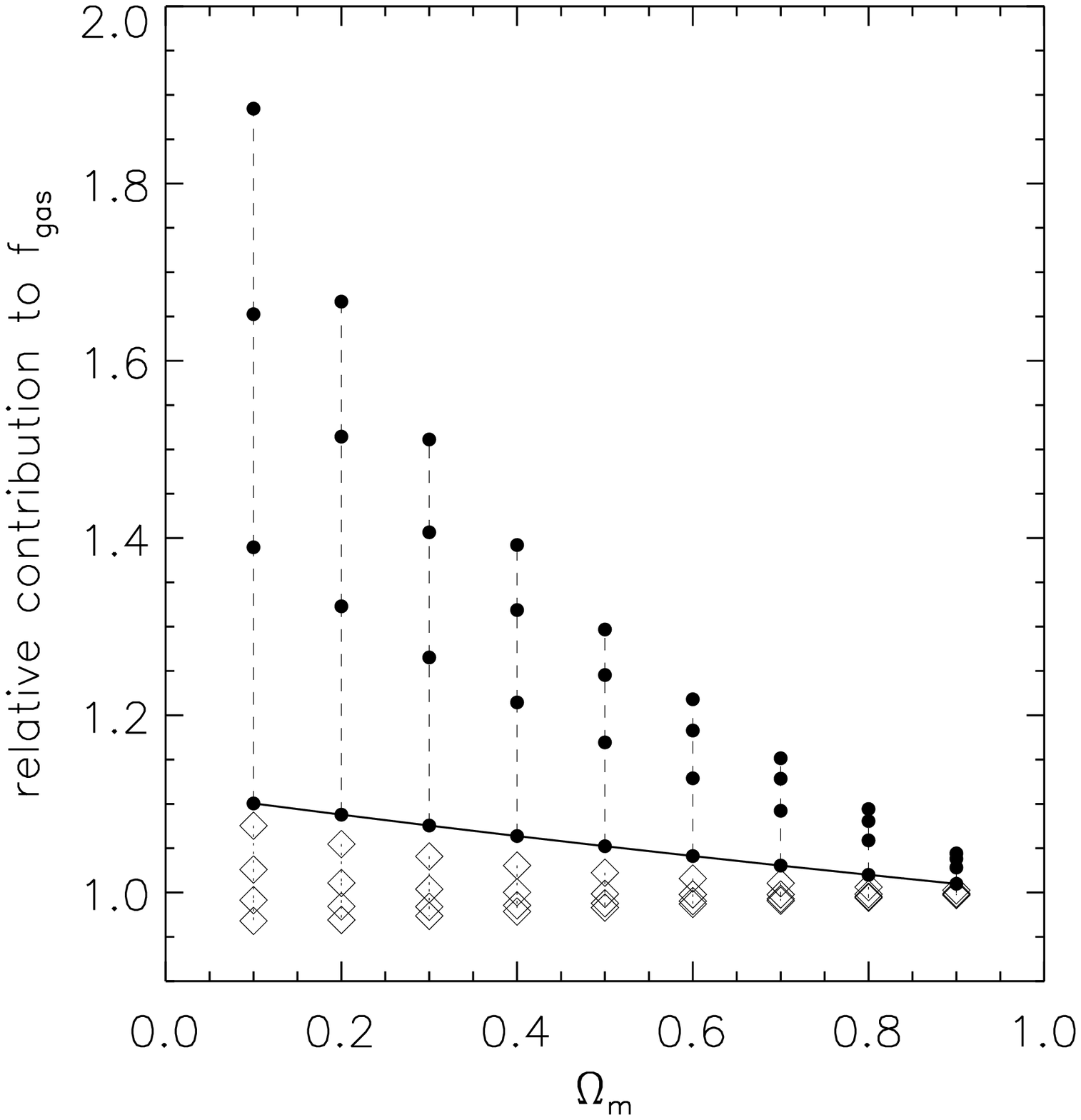,width=0.5\textwidth}  }
\caption{{\it (Left)} Values of the density contrast $\Delta$
(with respect to what expected for an Einstein -de Sitter model)
of the density contrast $\Delta$ estimated
in an ``$\Omega_{\rm m}+\Omega_{\Lambda}=1$''
universe at redshift 0, 0.1, 0.4, 0.7 and 1 ({\it filled circles},
from bottom to top, respectively) 
and with $w$ equal to --0.2, --0.6, --1
({\it diamonds}, from top to bottom, respectively;
{\it solid line}: values at $z=0$ for $w=-1$).
{\it (Right)} Plot of the multiplicative factors that correct
the gas fraction with respect to the initial estimate computed within
a density contrast of 1500 in an Einstein -de Sitter universe:
corrections due to the dependence upon
$d_{\rm ang}^{3/2}$ ({\it filled circles}, at $z=$ 0.1, 0.4, 0.7 and 1,
starting from the bottom, respectively) and
the change due to the density contrast ({\it diamonds};
{\it solid line}: values at $z=0.1$).
} \label{fig:cosmo_cfr} \end{figure*}

\section{THE COSMOLOGICAL FRAMEWORK}

We refer to $\Omega_{\rm m}$ as the {\it total} matter 
density (i.e., the sum of the {\it cold} and {\it baryonic} component:
$\Omega_{\rm m} = \Omega_{\rm c} + \Omega_{\rm b}$)
in unity of the critical density, $\rho_{\rm c} = 3H_0^2/ 
(8\pi G)$, where $H_0 = 100 h$ km s$^{-1}$ Mpc$^{-1}$ 
is the Hubble constant and $G$ is the gravitational constant,
and to $\Omega_{\Lambda}$ as the constant energy density associated 
with the ``vacuum" (Carroll et al. 1992). We consider a generalization 
from this static, homogeneous energy component to 
a dynamical, spatially inhomogeneous form of energy 
with still a negative pressure, or ``quintessence'' 
(e.g. Turner \& White 1997, Caldwell et al. 1998).
We neglect the energy associated to the radiation
of the cosmic microwave background, $\Omega_{\rm r} \approx
5 \times 10^{-5}$, and any possible contributions 
from light neutrinos, $\Omega_{\nu} = (\sum \mu_{\nu} /h^2
/ 92.5 eV)$, that is expected to be less than $0.05$
for a total mass in neutrinos, $\sum \mu_{\nu}$,
lower than 2.5 eV (see, e.g., the recent constraints from 2dF galaxy
survey in Elgaroy et al. 2002 and from combined analysis of cosmological
datasets in Hannestad 2002).
Thus, we can write $\Omega_{\rm m} + \Omega_{\Lambda} +\Omega_{\rm k} = 1,$
where $\Omega_{\rm k}$ accounts for the curvature of space.

In this cosmological scenario, the angular diameter distance
can be written as (e.g. Carroll, Press \& Turner 1992, cf. eqn.~25)
\begin{eqnarray}
d_{\rm ang} = & \frac{c}{H_0 (1+z)} \frac{S(\omega)}{|\Omega_{\rm
k}|^{1/2}},   \nonumber  \\
\omega = & |\Omega_{\rm k}|^{1/2} \int^z_0 \frac{d \zeta}{E(\zeta)},
\label{eq:dang}
\end{eqnarray}
where $S(\omega)$ is sinh$(\omega)$, $\omega$, $\sin(\omega)$ for
$\Omega_{\rm k}$ greater than, equal to and less than 0, respectively, and
\begin{eqnarray}
E(z) = \left[\Omega_{\rm m} (1+z)^3 +\Omega_{\rm k} (1+z)^2 +
\Omega_{\Lambda} (1+z)^{3+3 w}\right]^{1/2},
% \left[
% (1+\zeta)^2 (1+\Omega_{\rm m} \zeta) -\zeta(2+\zeta)\Omega_{\Lambda}
% \Omega_{\rm m} (1+\zeta)^3 +\Omega_{\rm k} (1+\zeta)^2 +
% \Omega_{\Lambda} (1+\zeta)^{3+3 w}
% \right]^{1/2}},
% \end{eqnarray}
\label{eq:ez}
\end{eqnarray}
that includes the dependence upon the ratio $w$
between the pressure and the energy density in
the equation of state of the dark energy component 
(Caldwell, Dave \& Steinhardt 1998, Wang \& Steinhardt 1998).
Hereafter we consider a pressure-to-density ratio
$w$ constant in time 
(see, e.g., Huterer \& Turner 2001, Gerke \& Efstathiou 2002
for the extension of eqn.~\ref{eq:ez} to a redshift-dependent form).
In particular, the case for a cosmological constant $\Lambda$ 
requires $w=-1$.

\subsection{The cosmological dependence of the observed 
Cluster Gas Mass Fraction}

We assume that galaxy clusters are spherically symmetric gravitationally
bound systems. 
For each galaxy cluster observed at redshift $z$,
we evaluate the gas mass fraction at $r_{\Delta}$, 
$f_{\rm gas}(r_{\Delta}) = M_{\rm gas}(<r_{\Delta}) /
M_{\rm tot}(<r_{\Delta})$, where $r_{\Delta}$
is defined according to the dark matter profile,
$M_{\rm tot}(<r)$, for a fixed density contrast
$\Delta = M_{\rm tot}(<r_{\Delta}) / (4\pi \rho_{\rm c, z} r^3_{\Delta})$.
In the latter equation, 
$\rho_{\rm c, z}$ is the critical density at redshift $z$
and is equal to $3 H_z^2/ (8 \pi G)$ with 
$H_z = H_0 E(z)$ (see eqn.~\ref{eq:ez}).
In the following sections, we describe how the dark matter mass profile
is obtained for each object and which density contrast
we adopt initially in an Einstein-de Sitter universe.

The assumed cosmological model affects 
the definition of the gas mass fraction, 
$f_{\rm gas}(r_{\Delta})$, given above in two independent ways:

\begin{enumerate}
\item for a galaxy cluster observed at redshift $z$ up to
a characteristic angular radius $\theta_{\rm c}$ and with
an X-ray flux $S_X = L_X (1+z)^{-4} / (4 \pi d_{\rm ang}^2) \propto
M_{\rm gas}^2 \theta_{\rm c}^{-3} d_{\rm ang}^{-3} / d_{\rm ang}^2$
[where the X-ray luminosity $L_X \approx n_{\rm gas}^2 \Lambda(T_{\rm gas})
\times \theta_{\rm c}^3 d_{\rm ang}^3$, 
and $\Lambda(T_{\rm gas})$ is the cooling
function of the X-ray emitting plasma that depends only on the plasma
temperature] and a total mass, $M_{\rm tot}$, estimated through the 
equation of the hydrostatic equilibrium,
the measured gas mass fraction is
\begin{equation}
f_{\rm gas} = \frac{M_{\rm gas}}{M_{\rm tot}}
\propto \frac{S_X^{1/2} \theta_{\rm c}^{3/2} 
d_{\rm ang}^{5/2}}{\theta_{\rm c} d_{\rm ang}}
\propto d_{\rm ang}(z, \Omega_{\rm m}, \Omega_{\Lambda}, w)^{3/2},
\label{eq:fgas_dang}
\end{equation}

\item the density contrast, $\Delta$, depends upon the redshift and the
cosmological parameters.
\end{enumerate}

We have initially evaluated the gas fraction,
$f_{\rm gas}(\Delta_{\Omega_{\rm m}=1})$, in an Einstein-de Sitter
universe with a Hubble constant of 50 km s$^{-1}$ Mpc$^{-1}$.
Then, we change the set of cosmological parameters and evaluate
for each cluster at redshift $z$ 
the new values of the angular diameter distance $d_{\rm ang}$
and the density contrast
with respect to the critical density at that redshift.
This density contrast is calculated according to the
formula in Lokas \& Hoffman (2001, Sect.~4.2)
for a ``$\Omega_{\rm m}+\Omega_{\Lambda}+\Omega_k=1$'' universe
and in Wang \& Steinhardt (1998, eqn.~5, 7 and A11; 
note that these formula are estimated for a background density 
that is $\Omega_{\rm m}(z)$ times the critical density) for
a ``quintessence'' flat model (see left panel in Fig.~\ref{fig:cosmo_cfr}).

Since we want to consider the gas fraction in each cluster
estimated within the same $\Delta$ for any given set of 
cosmological parameters, we therefore multiply
$f_{\rm gas}(\Delta_{\Omega_{\rm m}=1})$ by two factors,
the first one that rescales the distance, 
$F_1= (d_{\rm ang, \Omega_{\rm m}, \Omega_{\Lambda}, w}
/d_{\rm ang, \Omega_{\rm m}=1})^{3/2}$, 
and the second one that corrects by the change in the 
density contrast, $F_2 = (\Delta_{\Omega_{\rm m}, \Omega_{\Lambda}, w}/
\Delta_{\rm {\Omega_{\rm m}=1}}) \times (H_z^2 d_{\rm ang}^2)_{\Omega_{\rm m}=1}
/(H_z^2 d_{\rm ang}^2)_{\Omega_{\rm m}, \Omega_{\Lambda}, w}$,
where in the latter factor $M_{\rm tot}(<r_{\Delta})$ 
and the angular radius $\theta_{\Delta} = r_{\Delta} / d_{\rm ang}$ are 
given (see right panel in Fig.~\ref{fig:cosmo_cfr}).
In particular, being $\Delta \propto r_{\Delta}^{-2}$ and $f_{\rm gas}
\propto r_{\Delta}^{0.2}$ (e.g. Ettori \& Fabian 1999a, and figure~13 in
Frenk et al. 1999), we conclude that 
\begin{equation}
f_{{\rm gas}, \Omega_{\rm m}, \Omega_{\Lambda}, w}
= f_{{\rm gas} ,\Omega_{\rm m}=1} \times F_1 \times F_2^{-0.1}.
\end{equation}

As shown in Fig.~\ref{fig:cosmo_cfr}, the second correction
affects the $f_{\rm gas}$ values by less than 10 per cent
and is marginal with respect to the cosmological effects due to
the dependence upon the angular diameter distance.
We apply both these corrections to evaluate each cluster gas fraction 
in the following analysis.

\section{FIRST CONSTRAINT: $\Omega_{\rm m}$ FROM THE GAS FRACTION VALUE}

In this section, we describe how the local estimate of gas mass
fraction provides a robust constraint on the cosmic matter density.
We make use of the further assumption of a constant gas fraction with 
redshift in the next section, where we consider a sample of galaxy
clusters with $z>0.7$. 

\begin{table}
%%
%%  in local_*.dat
%%
\caption{The local sample from \sax MECS observations.
The quoted values are obtained from the deprojection of the spectral results and
assuming a functional form of the total mass profile (see Ettori et al. 2002
for details). A Hubble constant of 50 km s$^{-1}$ Mpc$^{-1}$ is considered
in an Einstein-de Sitter universe.
}
\begin{tabular}{l c c c}
cluster & $z$ &  $T_{\rm mw}(r_{\Delta})$ &  $f_{\rm gas}(r_{\Delta})$ \\
  & & &  \\
\multicolumn{4}{c}{$\Delta=1500$}  \\
A85   &  0.0518  & $5.77 \pm 0.32$ & $0.121 \pm 0.008$  \\
A426  &  0.0183  & $7.31 \pm 0.16$ & $0.172 \pm 0.009$  \\
A1795 &  0.0632  & $5.53 \pm 0.27$ & $0.130 \pm 0.009$  \\
A2029 &  0.0767  & $7.68 \pm 0.46$ & $0.126 \pm 0.007$  \\
A2142 &  0.0899  & $8.47 \pm 0.46$ & $0.176 \pm 0.011$ \\
A2199 &  0.0309  & $4.53 \pm 0.21$ & $0.123 \pm 0.009$ \\
A3562 &  0.0483  & $4.82 \pm 0.64$ & $0.117 \pm 0.027$ \\
A3571 &  0.0391  & $5.91 \pm 0.33$ & $0.104 \pm 0.009$  \\
PKS0745&  0.1028  & $8.36 \pm 0.47$ & $0.143 \pm 0.009$  \\

  & & &  \\
\multicolumn{4}{c}{$\Delta=500$}  \\
A85   &  0.0518  & $4.84 \pm 0.27$ & $0.134 \pm 0.011$  \\
A426  &  0.0183  & $8.12 \pm 0.17$  & $0.235 \pm 0.015$  \\
A1795 &  0.0632  & $4.59 \pm 0.22$ & $0.122 \pm 0.013$  \\
A2029 &  0.0767  & $6.30 \pm 0.37$  & $0.142 \pm 0.011$  \\
A2142 &  0.0899  & $7.19 \pm 0.34$  & $0.203 \pm 0.018$  \\
A2199 &  0.0309  & $4.21 \pm 0.20$ & $0.183 \pm 0.014$ \\
A3571 &  0.0391  & $4.24 \pm 0.23$ & $0.132 \pm 0.017$ \\
PKS0745&  0.1028  & $8.81 \pm 0.50$  & $0.126 \pm 0.012$  \\
\end{tabular}

\label{tab:local}
\end{table}

The observational constraints on the  abundance of the light elements 
(e.g. D, $^3$He, $^4$He, $^7$Li) in the scenario of the 
primordial nucleosynthesis gives a direct 
measurement of the baryon density with
respect to the critical value, $\Omega_{\rm b}$.
Moreover, the BOOMERANG, MAXIMA-1 
and DASI experiments have recently shown 
that the second peak in the angular power spectrum of the 
cosmic microwave background provides a constraint
on $\Omega_{\rm b}$ completely consistent with the one
obtained from calculations on the primordial nucleosynthesis
(e.g. de Bernardis et al. 2002).

If the regions that collapse to form rich clusters maintain the same
ratio $\Omega_{\rm b}/ \Omega_{\rm c}$ as the rest of the Universe, a
measurement of the cluster baryon fraction and an estimate of
$\Omega_{\rm b}$ can then be used to constraint the ``cold", and more
relaxed, component of the total matter density.  This method alone can
not provide a reliable limit on the amount of the mass-energy
presents in the Universe as {\it hot} constituents (e.g. WIMPS, like
massive neutrino) or energy of the field (e.g. $\Omega_{\Lambda}$,
quintessence), as both do not cluster on scales below 50 Mpc.

X-ray observations show that the dominant component of the 
luminous baryons is the X-ray emitting gas that falls into the 
cluster dark matter halo. Therefore, the gas fraction alone provides 
a reasonable upper limit on $\Omega_{\rm c}$:
\begin{equation}
\Omega_{\rm c} < \ \frac{\Omega_{\rm b} }{f_{\rm gas}} \propto  h^{-1/2},
\label{eq:fgas}
\end{equation}
where the dependence of the ratio $\Omega_{\rm b}$/$f_{\rm gas}$ on 
the Hubble constant is factored out (White et al. 1993, 
White \& Fabian 1995, David, Jones \& Forman 1995, 
Evrard 1997, Ettori \& Fabian 1999a, Mohr, Mathiesen \& Evrard 1999,
Roussel, Sadat \& Blanchard 2000, Erdogdu, Ettori \& Lahav 2002,
Allen, Schmidt \& Fabian 2002).

To assess this limit, we use the gas mass fraction 
estimated in nearby massive galaxy clusters
selected to be relaxed, cooling-flow systems with mass-weighted
$T_{\rm gas} >$4 keV from the sample presented 
in Ettori, De Grandi \& Molendi (2002; cf. Table~\ref{tab:local}).
To date, this sample is the largest for
which the physical quantities (i.e. profiles of gas density, temperature,
luminosity, total mass, etc.) have all been derived simultaneously from
spatially-resolved spectroscopy of the same dataset (\sax observations,
in this case). Through the deprojection of the spectral results,
and assuming a functional form for the dark matter distribution
to be either a King (King 1962) or a Navarro, Frenk \& White (1997)
profile,
the gas and total mass profiles are recovered in a self-consistent way.
Hence, the density contrast, $\Delta = M_{\rm tot}(<r_{\Delta})
/ (4\pi \rho_{\rm c, z} r^3_{\Delta})$, and the gas fraction 
at $r_{\Delta}$, $f_{\rm gas}(r_{\Delta}) = M_{\rm gas}(<r_{\Delta}) /
M_{\rm tot}(<r_{\Delta})$, can be properly evaluated.

\begin{figure}
\begin{center}
\epsfig{figure=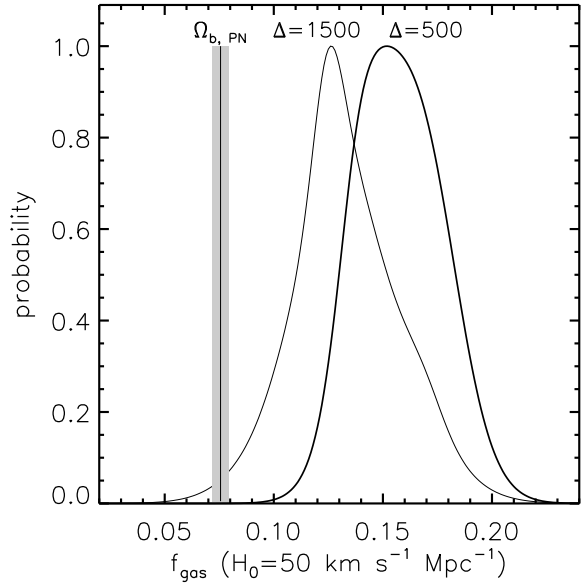,width=0.5\textwidth}
\caption{Probability distributions (see Press 1996)
at different density contrast $\Delta$ of the values of the
gas mass fraction for cooling-flow clusters with $T_{\rm gas} > 4$ keV 
from the sample in Ettori, De Grandi \& Molendi (2002). 
The central and $1-\sigma$ values are $(0.126, 0.024)$ and 
$(0.152, 0.020)$ at $\Delta=$ 1500 and 500, respectively.
These values have to be compared with the mean and standard deviation
of $(0.134, 0.025)$ and $(0.152, 0.043)$, respectively. 
The estimated $\Omega_{\rm b}$ from primordial nucleosynthesis results 
(Burles et al. 2001) is overplotted for comparison.
} \label{fgas_dc} \end{center} \end{figure}

\begin{figure}
\epsfig{figure=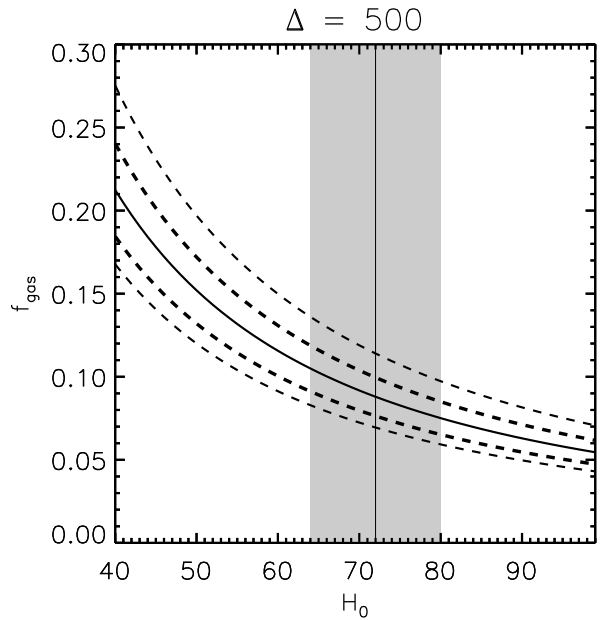,width=0.5\textwidth}
\caption{The observed gas fraction at $\Delta=500$ for the eight galaxy clusters
in Table~1 is here plotted as function of the Hubble constant
(shaded region: constraints from Freedman et al. 2001).
The solid line indicates the central value and the dashed lines
the 1 and 2 $\sigma$ uncertainties.
} \label{fig:hub} \end{figure}

In Fig.~\ref{fgas_dc}, we compare the estimated $\Omega_{\rm b}$
from primordial nucleosynthesis calculation with the probability
distribution (obtained following a Bayesian approach discussed
in Press 1996) of the values of the gas mass fraction for nine 
(at $\Delta=1500$) and eight (at $\Delta=500$) nearby massive objects.
This plot shows that at lower density contrast (i.e. larger radius)
the amount of gas mass tends to increase relatively to the underlying 
dark matter distribution. Even if there is indication that
the gas fraction becomes larger moving outward and within the
virialized region, we decide to adopt $f_{\rm gas}$ at 
$\Delta=$500 as representative of the cluster gas fraction.
In fact, at $\Delta=$500 the gas fraction is expected to be not 
more than 10 per cent less than the universal value, a
difference which is smaller than our statistical error, and in any
case is likely to be swamped by other effects 
(see comments in the first item in Section~5).

We plot in Fig.~\ref{fig:hub} the dependence upon the Hubble constant
of the $f_{\rm gas}$ value $(\propto h^{-3/2})$ estimated at $\Delta=$ 500
for the objects in the local sample.
Assuming $h = h_{\rm HST}= 0.72 \pm 0.08$ (from the results of the
HST Key Project on distances measured using Cepheid variables,
Freedman et al. 2001)
and $\Omega_{\rm b} h^2 = \Omega_{\rm b, PN} h^2 =
0.019 \pm 0.001$ (from primeval deuterium abundance and calculations on the
primordial nucleosynthesis, Burles, Nollett, Turner 2001),
we obtain that $\Omega_{\rm c}$ in eqn.~\ref{eq:fgas}
is less than $0.54$ (95.4 per cent confidence level).

\begin{figure*} \begin{center}
\hbox{
\epsfig{figure=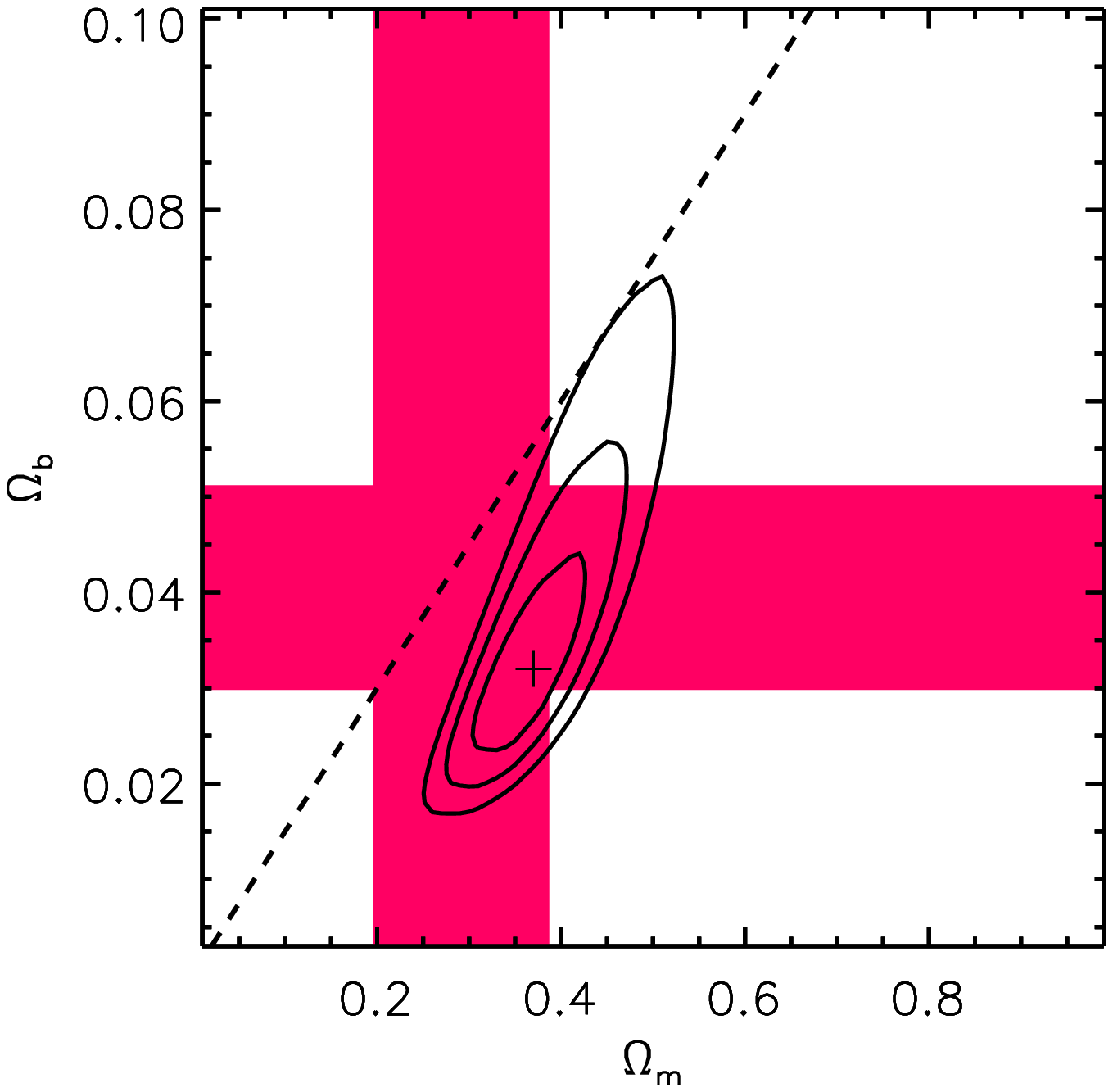,width=0.5\textwidth}
\epsfig{figure=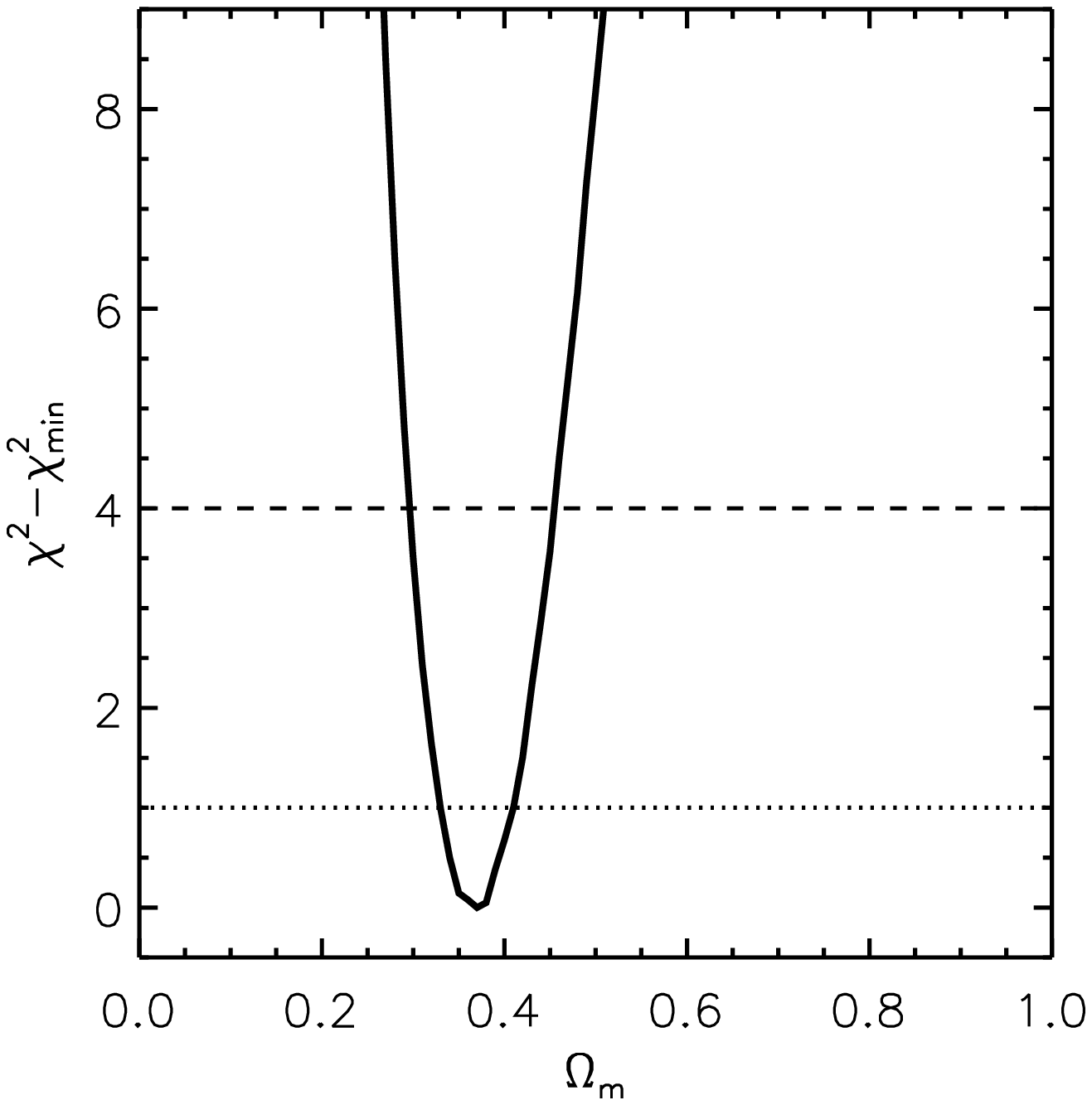,width=0.5\textwidth} }
\caption{ {\it (Left)}
Probability distribution contours ({\it solid lines}: $1, 2, 3 \sigma$ 
for two interesting parameters) in the
$\Omega_{\rm b}-\Omega_{\rm m}$ plane from marginalization of the likelihood
provided from the baryon fraction in clusters
assuming a $f_{\rm gal} = 0.02 (\pm 0.01) h^{-1}_{50}$
(White et al. 1993; Fukugita, Hogan \& Peebles 1998),
$\Omega_{\rm b, PN}$ (Burles et al. 2001), and
$H_0 = 72 \pm 8$ km s$^{-1}$ Mpc$^{-1}$ (Freedman et al. 2001).
As reference, $\Omega_{\rm b, CMB}$ (horizontal shaded region)
and $\Omega_{\rm m, CMB}$ (vertical shaded region; Netterfield et al. 2002),
and  $(\Omega_{\rm b}/\Omega_{\rm m})_{\rm 2dF}$ (dashed line
indicates the central value; Percival et al. 2001) are indicated.
% The dark shaded region represents the excluded region at
% $2 \sigma$ level (95.4 per cent) from the observed gas mass fraction
% in the sample of X-ray emitting galaxy clusters.
{\it (Right)}
Maximum likelihood distribution in the $\Omega_{\rm m}$ axis
after marginalization over the other parameters
(dotted line: $1 \sigma$, dashed line: $2 \sigma$).
} \label{ob_om} \end{center} \end{figure*}

% As stated above, this method constrains the clustered component of the
% dark matter, $\Omega_{\rm c}$. Any contribution from a {\it hot}
% component adds linearly and independently.
Including a contribution from stars in galaxies of about
$f_{\rm gal} =0.02 (\pm 0.01) h^{-1}_{50}$ (White et al. 1993,
Fukugita et al. 1998) and excluding any further components
to the baryon budget (see e.g. Ettori 2001), one can write
$\Omega_{\rm b} / \Omega_{\rm c} = f_{\rm gas} +f_{\rm gal} = f_{\rm b}$
and, consequently from the definition of $\Omega_{\rm m}$,
$\Omega_{\rm b} / \Omega_{\rm m} = f_{\rm b}/(1+f_{\rm b})$.
(Note that the estimate of $M_{\rm tot}$ does not include the contribution
of the gas mass, that would require the solution of a second order differential
equation instead of a much simpler, and usually adopted, 
first order equation of the spherical 
hydrostatic equilibrium).

Finally, we consider the eight relaxed nearby clusters $i$
with $T>4$ keV (see Table~\ref{tab:local}) to
evaluate the baryon fraction at redshift $z_i$, $f_{\rm b,i}$,
and within $\Delta(\Omega_{\rm m}=1, \Omega_{\Lambda}=0, w=-1) =$500.
For a given set of parameters $(\Omega_{\rm m}, \Omega_{\Lambda}, w)$
in the range $[0,1]$, $[0,2]$ and $[-1,0]$, respectively,
we estimate $f_{\rm b,i}$ (and its relative error $\epsilon_{\rm b,i}$
as propagation of the estimated error on $f_{\rm gas,i}$ and $f_{\rm gal}$,
where the error on $f_{\rm gas,i}$ comes from the measured uncertainties
on the gas and total mass estimates in Ettori et al. 2002, and
the error on $f_{\rm gal}$ is $0.01 h_{50}^{-1}$)
after considering the cosmological dependence of both $d_{\rm ang}$
and $\Delta(\Omega_{\rm m},\Omega_{\Lambda},w)$
(more relevant for high$-z$ systems, see Section~2) and
calculate
\begin{eqnarray}
\chi^2_A & = \sum_i \frac{ \left( \ f_{\rm b, i} -\Omega_{\rm b}/\Omega_{\rm m}
\ \right)^2}{\epsilon_{\rm b, i}^2} + \nonumber  \\
 & \frac{ \left(\Omega_{\rm b}-\Omega_{\rm b, PN} \right)^2}{\epsilon_{\Omega_{\rm b}}^2} +
\frac{ \left(h- h_{\rm HST}\right)^2}{\epsilon_h^2},
\label{chi2a}
\end{eqnarray} 
where $\Omega_{\rm b, PN}$ and $h_{\rm HST}$ are defined above.
% the dependences upon the redshifts are written as
% $h_z = h E(z)$ (see eqn.~\ref{eq:ez}) and
% \begin{equation}
% \Omega_{\rm m, z} = \frac{\rho_0 (1+z)^3}{\rho_{{\rm c}, z}} =
% \Omega_{\rm m, 0} \frac{(1+z)^3}{E(z)^2}
% \end{equation}
% for a dust-matter universe like a cold dark matter model.

The $\chi^2$-distribution in eqn.~\ref{chi2a} is used to
construct a $\Delta \chi^2$ statistics,
$\Delta \chi^2 = \chi^2_{A} - \chi^2_{A, {\rm min}}$,
by which we generate regions and intervals of confidence
(e.g., the $1 \sigma$ level of confidence for one and
two degrees of freedom is
$\Delta \chi^2 =$ 1 and 2.3, respectively. A $\chi^2_{A, {\rm min}}$ = 29.3
is obtained).
% We obtain a minimum $\chi^2_A$ with ten degrees of freedom 
% of 31.1.
% probability for a random variable following a
% $\chi^2$ distribution to be larger than this of 4 per cent.

% marginalizing over the interval
% $[0.02,0.14]$ for $\Omega_{\rm b}$ and $[0.4,2.4]$ for
% $h=H_0/(50$ km s$^{-1}$ Mpc$^{-1})$:
Marginalizing over the accepted ranges of Hubble constant
from the HST Key Project (Freedman et al. 2001) and $\Omega_{\rm b}$ 
from primordial nucleosynthesis, we obtain ($2 \sigma$)
$\Omega_{\rm m} = 0.37^{+0.07}_{-0.08}$ and  
$\Omega_{\rm b} = 0.032^{+0.017}_{-0.010}$,
that are well in agreement with $\Omega_{\rm m, CMB}$ from CMB and
$\Omega_{\rm b}/\Omega_{\rm m}$ from large scale structures 
analysis of the ``2dF'' data (see Figure~\ref{ob_om}).

% Integrating the Sheth \& Tormen (1999) mass function parametrized 
% through the results in Pierpaoli et al. (2001) over the mass range
% $10^{14}-10^{16} M_{\odot}$, we obtain that the
% contribution of the clusters mass to the total density is about
% 2.4 per cent of the critical value and
% $\Omega_{\rm b, cl} = \Omega_{\rm m, cl} \times f_{\rm bar}$
% is 0.0026 using the local gas mass fraction estimate. 
% In other words, galaxy clusters contain just about 8 per cent 
% of the cosmic baryons
% (see also Fukugita, Hogan \& Peebles 1998). 

\section{SECOND CONSTRAINT: GAS FRACTION CONSTANT IN TIME}

\begin{figure}
\epsfig{figure=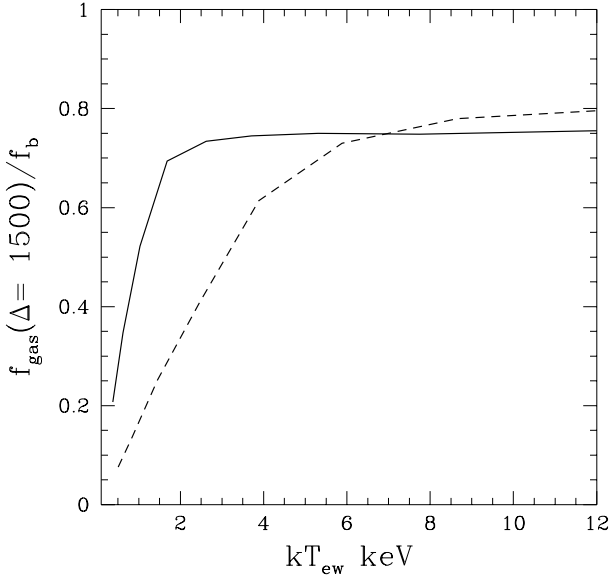,width=0.5\textwidth}
\caption{
The gas fraction in unit of the cosmic baryon budget
at $\Delta = 1500$ as a function of the observed
(emission weighted) temperature computed for the model of Tozzi \&
Norman (2001) with a constant entropy of 0.3 $\times 10^{34}$ erg cm$^2$ 
g$^{-5/3}$ in a
$\Lambda$CDM ($\Omega_{\rm m}= 1- \Omega_{\Lambda}=0.3$) Universe.
The {\it solid line} is for $z=0$, and the {\it dashed line} for $z=1$.
Note the 20 per cent offset from the universal value, which reduces
to 10 per cent for $\Delta = 500$ (see comments on the baryonic depletion
in Section~5).
} \label{fbz} \end{figure}

In this section, we present a sample of galaxy clusters with $z>0.7$
and compute their gas mass fraction. These values are compared
to the local estimates to put cosmological constraints 
under the assumption that the gas fraction
remains constant with redshift when computed at
the same density contrast.
It is worth noticing that this method,
originally proposed from
Sasaki (1996; see also Cooray 1998, Danos \& Ue-Li Pen 1998,
Rines et al. 1999, Ettori \& Fabian 1999b, Allen et al. 2002),
does not require any {\it prior} 
on the values of $\Omega_{\rm b}$ and $H_0$, taking into account 
just the relative variation of the gas fraction as function of time.
In other words, the method assumes that gas fraction in galaxy clusters
can be used like a ``standard candle'' to measure the 
geometry of the Universe.
This is a reasonable assumption in any hierarchical clustering
scenario when the energy of the ICM is dominated by the gravitational
heating and is supported by numerical
and semianalytical models for the thermodynamics of the ICM, also when
including preheating and cooling effects.  
In  recent hydrodynamical
simulations with an entropy level of 50 keV cm$^{-2}$ generated in the
cluster at $z=3$ (see Borgani et al. 2002), the baryon fraction in
clusters with an observed temperature around 3 keV is constant in time
within few percent (see also figure 13 in Bialek, Evrard and Mohr 2001).
Similar results are obtained in the semianalytical model of Tozzi \&
Norman (2001), where a constant entropy floor is initially present in
the cosmic baryons.  In Figure~\ref{fbz}, we show the prediction for the
baryonic fraction (in terms of the universal value)  within an average
overdensity $\Delta = 1500$ as a function of the emission weighted
temperature $T_{ew}$, computed in a $\Lambda$CDM cosmology and with an
entropy level of 0.3 $\times 10^{34}$ erg cm$^2$
g$^{-5/3}$ (see Tozzi \& Norman 2001 for details).  
A significant decrease of the baryonic fraction is expected at $z=1$
(dashed line)  with respect to the local value (solid line) only for
temperatures below 4 keV.  In particular, at temperatures above 6 keV,
the baryonic fraction is constant or possibly 5 per cent higher at $z=1$ with
respect to $z=0$.  This strongly supports the assumption of a baryon
fraction constant with the cosmic epoch for clusters with $kT>4$ keV.

With this assumption, we expect to measure a constant average gas
fraction locally and in distant clusters. However, the gas 
fraction is given by a combination of the observed flux and 
of the angular distance, and thus it depends on cosmology
(see discussion in Section~2). As shown in Figure~\ref{fig:cosmo_cfr},
the high redshift objects are more affected from this dependence and
show lower $f_{\rm gas}$ with respect to the local values
when universes with high matter density are assumed. 
By requiring the measured gas fractions to be constant as
a function of redshift, one can constrain the range of values of 
cosmological parameters which satisfies such a condition. 

\begin{table*}   
%% fext='_fgas.inp' & filefgas ='../cl1113/cl1113'+fext
%% fgas_fit, filefgas, [0, 50,1.,0.,-1], fgas, res_out
%% fgas_fit_info, filefgas, res_out
\caption{The high redshift sample from Chandra observations.
The results are obtained applying a $\beta-$model in an Einstein-de Sitter universe
with a Hubble constant of 50 km s$^{-1}$ Mpc$^{-1}$.
All the quoted errors are at $1 \sigma$ level.
Note that MS1054.5-0321 presents significant substructure (e.g. Jeltema et al. 2001).
The temperature and the best-fit of the surface brightness profile are estimated 
from the main body of the cluster once a circular region centered at 
(RA, Dec; 2000)=$(10^{\rm h} 56^{\rm m} 55"7,-3^o 37' 37'')$ 
and with radius of 36 arcsec is masked.
}
\begin{tabular}{l c c c c c c c c c} 
cluster & $z$ &  $r_{\rm out}$ & $T_{\rm gas}$ &  $r_{\rm c}$ & $\beta$ & $n_{\rm ele}(0)$ &
  $M_{\rm tot} (r_{\rm out})$ & $\Delta (r_{\rm out})$ & $f_{\rm gas} (r_{\rm out})$ \\
    & & '' / kpc & keV & kpc & & $10^{-2}$ cm$^{-3}$  & $10^{13} M_{\odot}$ & & \\
  & & & & & & & & & \\
%% 'lynx/cl1'
%RXJ0848+4453  &  1.273 & 
%  29.5 / 254 & $2.9_{-0.9}^{+1.6}$ & $219^{+205}_{-124}$ & $0.61^{+0.79}_{-0.26}$ & 
%  $0.31^{+0.13}_{-0.07}$ & $2.8^{+1.6}_{-1.0}$ & $503^{+283}_{-172}$ & $0.123^{+0.101}_{-0.053}$ \\
% 'lynx/cl3'
RDCS J0849+4452  &  1.261 & 
  29.5 / 254 & $5.0_{-1.0}^{+1.4}$ & $97^{+56}_{-31}$ & $0.68^{+0.28}_{-0.13}$ & 
  $0.96^{+0.24}_{-0.19}$ & $8.3^{+2.9}_{-2.1}$ & $1506^{+529}_{-376}$ & $0.048^{+0.016}_{-0.012}$ \\
% 'cl0910/cl0910'
RDCS J0910+5422  &  1.101  & 
  29.5 / 253 & $5.0_{-1.0}^{+1.3}$ & $114^{+61}_{-42}$ & $0.65^{+0.33}_{-0.17}$ & 
  $0.81^{+0.19}_{-0.13}$ & $7.9^{+2.6}_{-2.0}$ & $1802^{+583}_{-460}$ & $0.046^{+0.015}_{-0.010}$ \\
% 'ms1054/ms1054'
MS1054.5-0321 &  0.833  & 
  82.7 / 685 & $10.1_{-0.9}^{+1.1}$ & $576^{+51}_{-43}$ & $1.36^{+0.16}_{-0.12}$ & 
  $0.54^{+0.01}_{-0.01}$ & $61.3^{+7.2}_{-6.0}$ & $1064^{+125}_{-103}$ & $0.109^{+0.011}_{-0.011}$ \\
% 'rxj1716/rxj1716'
NEP J1716.9+6708 & 0.813 & 
  33.5 / 276 & $7.1_{-0.8}^{+1.0}$ & $116^{+14}_{-13}$ & $0.60^{+0.03}_{-0.03}$ & 
  $1.36^{+0.12}_{-0.11}$ & $11.1^{+1.6}_{-1.3}$ & $3048^{+433}_{-360}$ & $0.080^{+0.012}_{-0.011}$ \\
% 'rxj1350/rxj1350'
RDCS J1350.0+6007 & 0.804 &
  68.9 / 567 & $4.1_{-0.6}^{+0.8}$ & $191^{+61}_{-43}$ & $0.57^{+0.13}_{-0.08}$ & 
  $0.47^{+0.06}_{-0.06}$ & $13.2^{+3.3}_{-2.5}$ & $425^{+106}_{-81}$ & $0.192^{+0.041}_{-0.036}$ \\
% 'ms1137/ms1137'
MS1137.5+6625  & 0.782  & 
  45.3 / 370 & $6.3_{-0.4}^{+0.4}$ & $116^{+7}_{-7}$ & $0.67^{+0.02}_{-0.02}$ & 
  $1.69^{+0.08}_{-0.07}$ & $15.9^{+1.1}_{-1.0}$ & $1895^{+132}_{-125}$ & $0.104^{+0.008}_{-0.008}$ \\
% 'cl1113/cl1113'
WARPS J1113.1-2615 & 0.730  & 
  51.2 / 412 & $5.0_{-0.7}^{+0.8}$ & $109^{+27}_{-22}$ & $0.65^{+0.12}_{-0.09}$ & 
  $0.97^{+0.11}_{-0.09}$ & $14.1^{+3.1}_{-2.5}$ & $1336^{+297}_{-239}$ & $0.063^{+0.012}_{-0.012}$ \\
% 'cl2302/cl2302'
RDCS J2302.8+0844 & 0.720 &
  49.2 / 394 & $6.7_{-0.9}^{+1.1}$ & $118^{+24}_{-16}$ & $0.57^{+0.07}_{-0.05}$ & 
  $0.75^{+0.06}_{-0.06}$ & $15.3^{+3.0}_{-2.5}$ & $1681^{+332}_{-273}$ & $0.075^{+0.013}_{-0.012}$ \\
\end{tabular}

\label{tab:highz}
\end{table*}

\subsection{The high$-z$ sample}

\begin{figure}
\psfig{figure=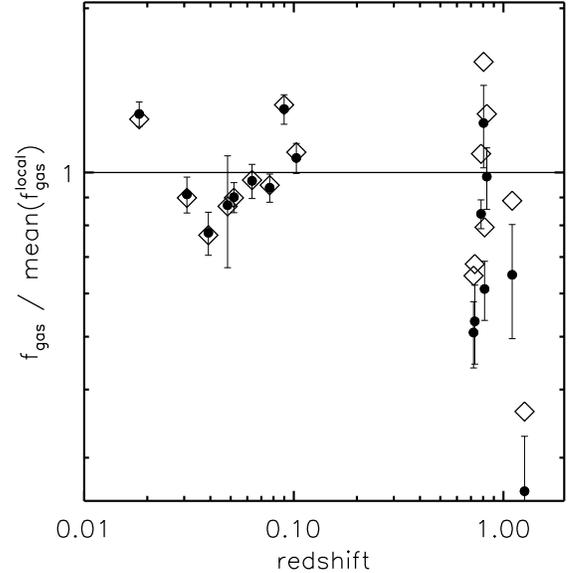,width=0.5\textwidth}
\caption{
Distribution as function of redshift of the gas fraction relative to the mean
local value and estimated for the clusters in our sample at the same
overdensity $\Delta=$1500.
{\it Filled circles} are the values calculated in an Einstein-de Sitter universe,
whereas {\it diamonds} indicate the results for a
low density universe ($\Omega_{\rm m} = 1- \Omega_{\Lambda}=0.3$). 
} \label{fig:fgas_z} \end{figure}

We define a local sample considering all the relaxed systems
in Ettori et al. (2002) that have a mass-weighted temperature 
larger than 4 keV within $\Delta=1500$ (cf. Table~\ref{tab:local}).
Within this density contrast, we measure $f_{\rm gas}$ as described in Section~2.
The density contrast is chosen 
to be $1500$ as a good compromize between the cluster regions directly 
observed in the nearby systems and those with relevant X-ray emission 
in the high$-z$ objects.
This sample includes eight high redshift ($z > 0.7$)
hot ($T > 4$ keV) clusters, four of which selected from the \rosat Deep Cluster
Survey (RDCS; Rosati et al. 1998, Stanford et al. 2001, Holden et al. 2002), 
two from the {\it Einstein} Extended Medium Sensitivity Survey (MS; Gioia et al. 1990), 
one from the Wide Angle \rosat Pointed Survey (WARPS; Perlman et al. 2002,
Maughan et al. 2002) and one part of the North Ecliptic Pole survey 
(NEP; Gioia et al. 1999, Henry et al. 2001).
We reprocess the level=1 events files retrieved 
from the archive and obtain a spectrum and an image for each
cluster (see details in Tozzi et al., in preparation).
Seven of these objects were observed in ACIS-I mode (only MS1054 has
been observed with the back-illuminated S3 CCD).
Following the prescription in Markevitch \& Vikhlinin (2001),
the effective area below 1.8 keV in the front-illuminated CCD
is corrected by a factor 0.93 to improve the cross calibration
with back-illuminated CCDs.
The spectrum extracted up to the radius $r_{\rm out}$ to
optimize the signal-to-noise ratio
is modelled between 0.8 and 7 keV with an absorbed optically--thin plasma
({\tt wabs(mekal)} in XSPEC v.~11.1.0, Arnaud 1996)
with fixed redshift, galactic absorption 
(from radio HI maps in Dickey \& Lockman 1990) and metallicity
(0.3 times the solar values in Anders \& Grevesse 1989) and using
a local background obtained from regions of the same CCD free of any
point source.
The gas temperature and the normalization
$K$ of the thermal component are the only free parameters.
The surface brightness profile obtained from the image is fitted
with an isothermal $\beta-$model (Cavaliere \& Fusco-Femiano 1976, 
Ettori 2000), which provides an analytic expression for the
gas density and total mass of the cluster. 
In particular, the central electron density is obtained
from the combination of the best-fit results from the spectral and
imaging analyses as follows:
\begin{equation}
n^2_{0, \rm ele} = \frac{4 \pi d_{\rm ang}^2 \times (1+z)^2 \times K \times 10^{14}}
% {0.82 \times 4 \pi r_{\rm c}^3 \int_0^{r_{\rm out}} (1+x^2)^{-3\beta} x^2 dx}
{0.82 \times 4 \pi r_{\rm c}^3 \times EI},
\end{equation}
where the emission integral is estimated by integrating 
along the line of sight the emission from the spherical source
up to 10 Mpc, $EI = \int_0^{x_1} (1+x^2)^{-3\beta} x^2 dx
+ \int_{x_1}^{x_2} (1+x^2)^{-3\beta} x^2 (1-\cos \theta) dx$,  with
$\theta = \arcsin(r_{\rm out}/r)$, $x_1 = r_{\rm out}/r_{\rm c}$ and
$x_2 = 10 {\rm Mpc} /r_{\rm c}$,
$(\beta, r_{\rm c})$ are the best-fit parameters of the $\beta-$model
and we assume $n_{\rm p} = 0.82 n_{\rm e}$ in the ionized intra-cluster plasma.
After 1000 random selections of a temperature, normalization
and surface brightness profile (drawn from Gaussian distributions with mean and variance 
in accordance to the best-fit results),
we obtain a distribution of the estimates
of the gas and total mass and of the gas mass fraction.
We adopt for each cluster the median value and the 16th and 84th percentile
as central and $1 \sigma$ value, respectively (see Table~\ref{tab:highz}).

In Figure~\ref{fig:fgas_z}, we plot $f_{\rm gas}$ of the clusters 
in exam as a function of redshift, as computed in an Einstein-de Sitter
(filled dots) and a low density ($\Omega_{\rm m}=1- \Omega_{\Lambda}=0.3$)
universe (open squares).

\subsection{The analysis}
We compare our local estimate of the gas mass fraction 
with the values observed in the objects at $z>0.7$ within the same
density contrast 
$\Delta = M_{\rm tot}(<r_{\Delta}) / (4\pi \rho_{\rm c, z} r^3_{\Delta})$,
where $M_{\rm tot}(<r_{\Delta})$ is estimated from the
$\beta-$model.
Initially, we estimate the gas fraction for all the clusters 
at $\Delta=$ 1500 in an Einstein-de Sitter universe.
Then, we proceed as described in Section~2.

\begin{figure}
\epsfig{figure=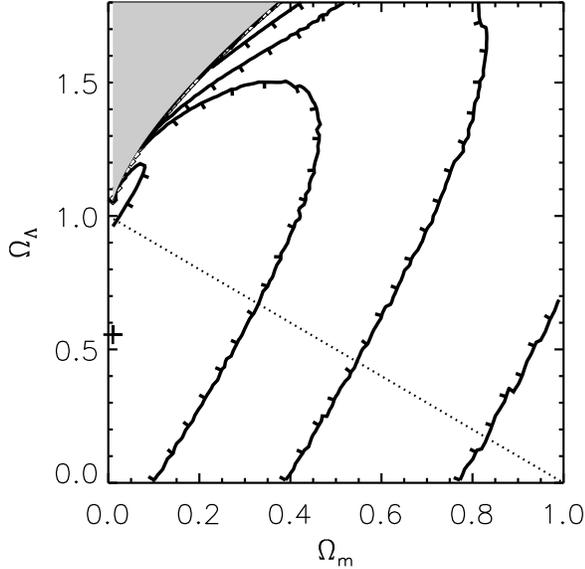,width=0.5\textwidth}
\caption{Maximum likelihood distributions in the ``$\Omega_{\rm m}+
\Omega_{\Lambda}+\Omega_k=1$" region obtained from the application of
the method that requires ``$f_{\rm gas}$=constant"
(cf. Section~4; shaded region: no-big-bang solution, dotted line:
``$\Omega_k=0$'' region). 
The contours enclose the regions with $\Delta \chi^2=$2.30, 6.17, 11.8
(with respect to the minimum of 11.7 with eight objects in exam),
corresponding to 1, 2, 3 $\sigma$, respectively, for a distribution
with two interesting parameters.
} \label{fig:dang} \end{figure}

As discussed above, we consider a set of parameters 
$(\Omega_{\rm m}, \Omega_{\Lambda})$ in the range
$[0,1]$ and $[0,2]$, respectively,
both fixing $w$ equals to $-1$ as prescribed for the ``cosmological constant''
case and exploring the range $w \in [-1, 0]$ 
(in this case, we require $\Omega_k=0$),
and evaluate the distribution
\begin{equation}
\chi^2_B = \sum_j \frac{ \left[ \ f_{\rm gas, j}(\Omega_{\rm m},\Omega_{\Lambda}) -
\overline{f_{\rm gas}} \ \right]^2}{\epsilon_{\rm gas, j}^2 +\overline{\epsilon_{\rm gas}}^2},
\label{chi2b}
\end{equation}
where $\overline{f_{\rm gas}}$ and $\overline{\epsilon_{\rm gas}}$
are the mean and the standard deviation of the values 
of the gas fraction in the {\it local} cluster sample,
and $\epsilon_{\rm gas, j}$ is the error on the measurement of
$f_{\rm gas, j}$ for $j \in [$high--z sample$]$.
It is worth noticing that the use of the standard deviation around
the mean is a conservative approach. For example, at ($\Delta, 
\Omega_{\rm m}, H_0) = (1500, 1, 50)$, we measure a mean of $0.134$ and
standard deviation of $0.025$, the latter being about 8 times 
larger than the measured error on the weighted mean and slightly
larger than the dispersion obtained with the Bayesian method 
illustrated in Fig.~\ref{fgas_dc}.
Using just the results obtained on the $\chi_B^2$ distribution, 
we obtain a best-fit solution that requires 
the following upper limits at $2 \sigma$ level (one interesting parameter),
$\Omega_{\rm m}<0.64$ and $\Omega_{\Lambda}<1.69$
(cf. Figure~\ref{fig:dang}).

\section{SYSTEMATIC UNCERTAINTIES}
 
The intrinsic scatter in the distribution of 
the $f_{\rm gas}$ values of approximately
20 per cent is the most relevant statistical uncertainty affecting
our cosmological constraints.   We discuss in this section how
a number of systematic uncertainties contribute to a lesser extent.

\begin{itemize}
\item We have assumed that the intracluster medium is in hydrostatic equilibrium,
distributed with a spherical geometry, with no significant (i) clumpiness in the X-ray
emitting plasma, (ii) depletion of the cosmic baryon budget at the reference radius,
and (iii) contribution from non-thermal components.
The hydrostatic equilibrium in a spherical potential is widely
verified to be a correct assumption for local clusters, but it cannot 
be the case
(in particular on the geometry of the plasma distribution, 
but see Buote \& Canizares 1996, Piffaretti, Jetzer \& Schindler 2002) 
for high redshift clusters.
The level of clumpiness in the plasma expected from numerical simulations
(e.g. Mohr et al. 1999) induce an overestimate of the gas fraction.
On the other hand, the expected baryonic depletion 
(still from simulations; e.g. Frenk et al. 1999) underestimates  
the observed cluster baryon budget by an amount that can be 
comparable (but of opposite sign) to the effect of 
the clumpiness (see discussion in Ettori 2001).
For the high redshift objects observed within $\Delta$=1500, 
a larger clumpiness can be present as consequence
of on-going process of formation that, on the other hand, 
might still partially compensate for a residual depletions of baryons 
observed in the simulations in the range 4--6 keV (cfr. Figure~\ref{fbz}).
The combination of these effects induce however an uncertainty 
in the relative amount of baryons of less than 10 per cent, 
percentage that is below the observed statistical uncertainties.

Moreover, the radial dependence of $f_{\rm gas}$ can be slightly steeper
in the inner cluster regions due to the relative broader distribution 
of the gas with respect to the dark matter
(but see Allen et al. 2002 for $f_{\rm gas}$ profiles observed
flat within $\Delta$=2500).
Finally, from the remarkable match between the gravitational mass profiles obtained
independently from X-ray and lensing analysis (e.g. Allen, Schmidt \& Fabian 2001),
it seems negligible any non-thermal contribution to the total mass apart
from a possible role played in the inner regions.
 
\item The high redshift sample has been analyzed using an isothermal $\beta-$model.
From recent analyses of nearby clusters with spatially resolved temperature profiles, 
it appears that this method can still provide a reasonable constraint on the
gas density, but surely affect the estimate of the total mass, in particular
in the outskirts and, more dramatically, when an extrapolation is performed.
On the other hand, the physical properties of nearby clusters 
at overdensity of 1500 (e.g. Ettori et al. 2002) guarantee that the expected
gradient in temperature is not significantly steep and, therefore, does not affect
the estimate of the total gravitating mass.
However, the presence of a gradient in the temperature profile would reduce
the total mass measurements and increase, consequently, the derived
gas mass fraction.

\item We assume that relaxed nearby and high-$z$ clusters, 
both with $T_{\rm gas} > 4$ keV, represent a homogeneous class of objects.
If we relax this assumption and include non-relaxed nearby systems
(or equivalently no-cooling-flow clusters in the sample of Ettori et al. 2002),
we obtain the same cosmological constraints though the scatter in the
distribution of the gas fraction values becomes larger
(e.g., at $\Delta=$1500, the scatter is 0.025 for relaxed systems only and
0.033 for the complete sample).
\end{itemize}

In conclusion, we check our results against two effects: 
(i) by increasing the gas mass fraction in the high$-z$ sample by a factor 1.15,
and (ii) by changing the slope of the radial dependence of $f_{\rm gas}(r)$ 
(cfr. factor $F_2$ in Section~2) by $\pm 0.1$.
All these corrections do not change significantly
the results quoted in the next section on $\Omega_{\rm m}$
due to heavier statistical weight given from the first method, that is the 
one less (or completely not) affected from the above mentioned effects.
On the other hand, they mostly affect the second method in such a way that 
(i) gas fraction values at high$-z$ higher by 15 per cent 
reduce the $2 \sigma$ upper limit on $\Omega_{\Lambda}$ by 10 per cent, 
but increase the one on $w$ by 50 per cent, 
(ii) steeper radial $f_{\rm gas}$ profiles (i.e. larger dependence upon
the density contrast $\Delta$) raise the limit on $\Omega_{\Lambda}$, 
but have not relevant effects on $w$.
In details, when the slope changes from 0 to $-0.1$ and $-0.2$,
the upper limit on $\Omega_{\Lambda}$ increases from 25 to 5 per cent, 
respectively. 

\begin{figure*}
\hbox{
\epsfig{figure=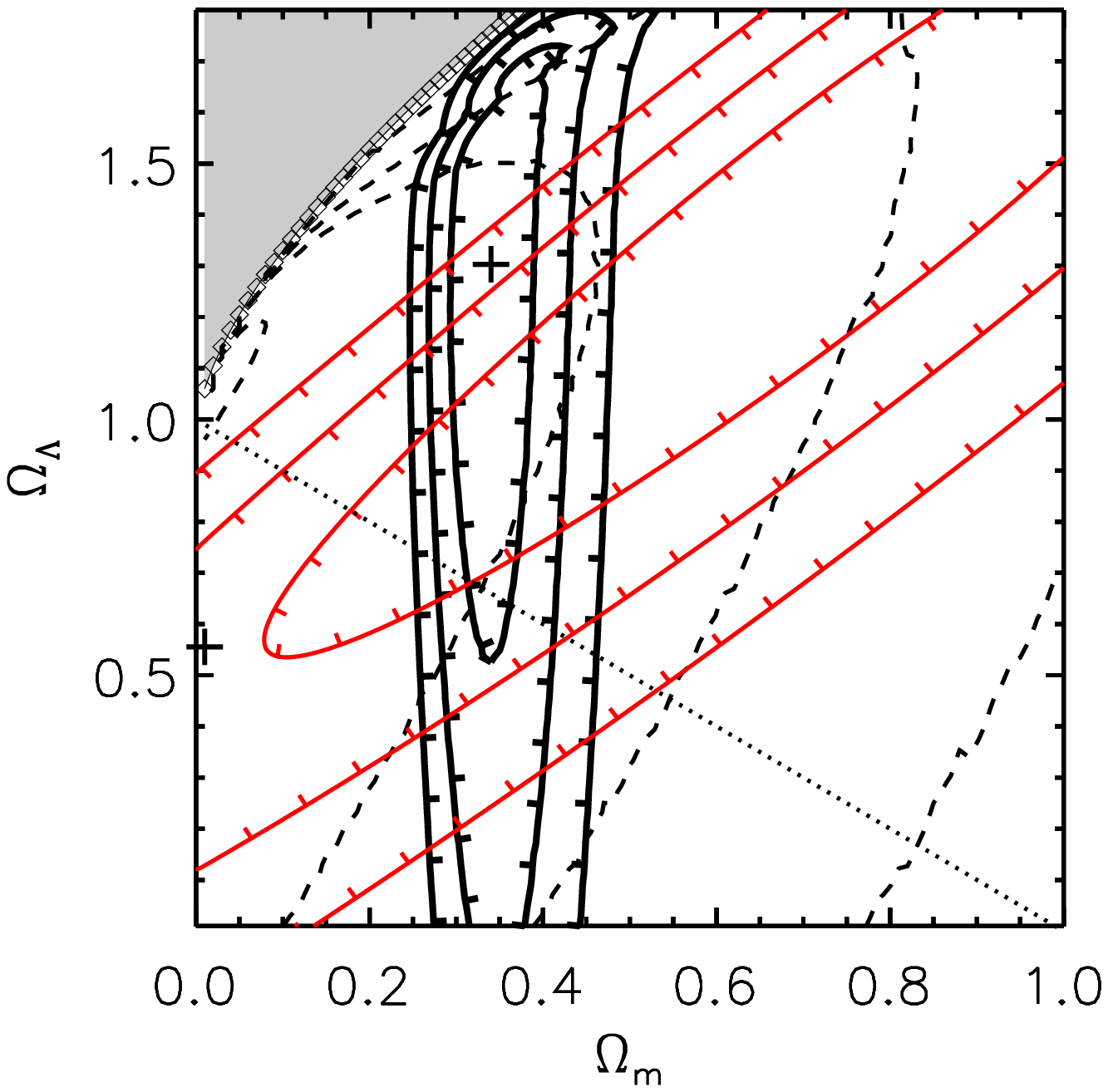,width=0.5\textwidth}
\epsfig{figure=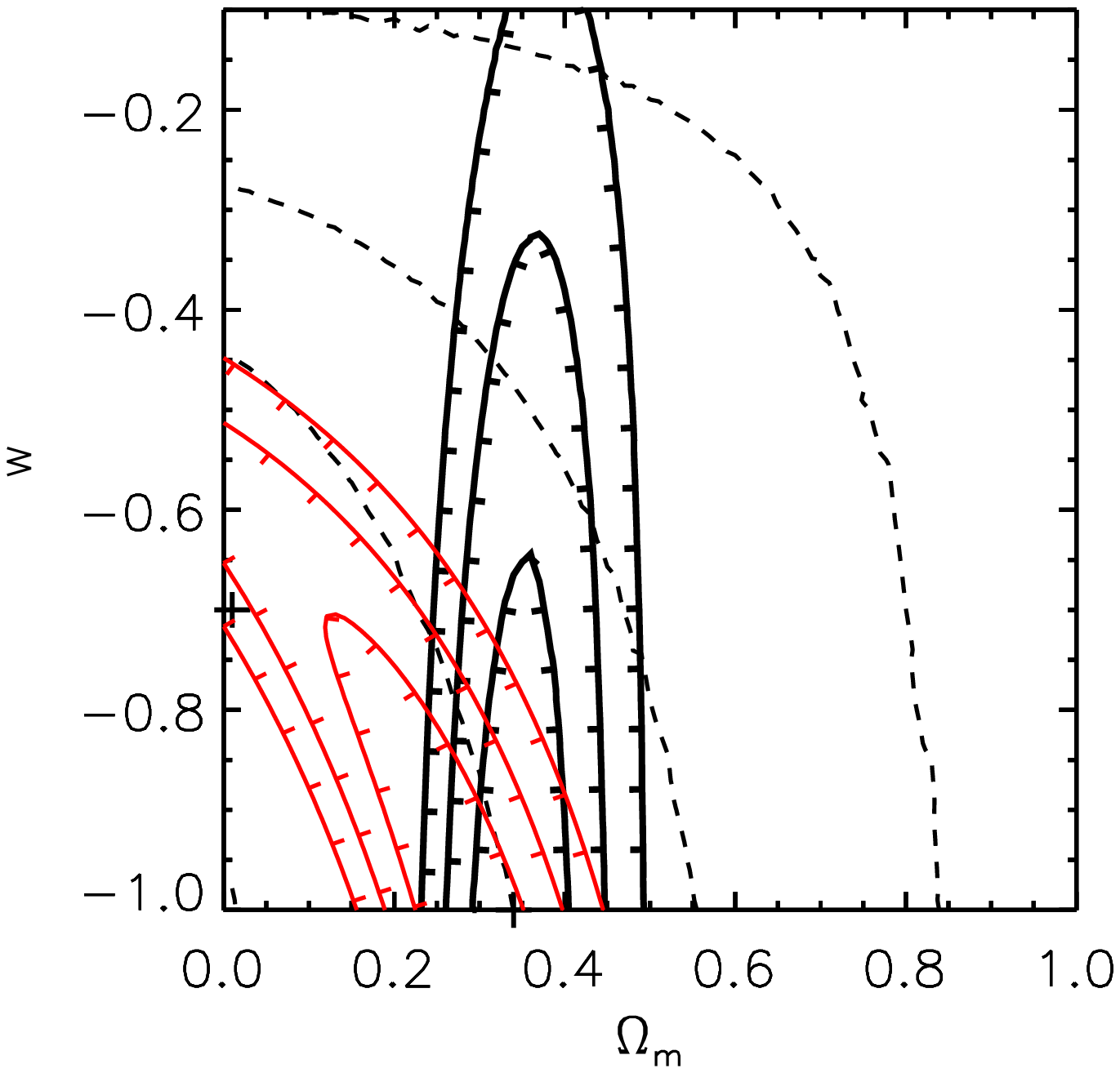,width=0.5\textwidth}
}
\caption{
{\it (Left)} Maximum likelihood distributions in the ``$\Omega_{\rm m}+
\Omega_{\Lambda}+\Omega_k=1$" region.
Contour plots (thick solid lines) from the combination of the
two likelihood distributions ($A$: cluster baryonic content, $B$: gas fraction constant
with redshift; dashed lines indicate the constraints from the second method only, 
see Fig.~\ref{fig:dang}; the cross indicates the best-fit result at 
$\Omega_{\rm m}=0.34$, $\Omega_{\Lambda}=1.30$) with overplotted the
constraints from the magnitude-redshift method applied to a set of SN Ia
(cf. Leibundgut 2001, thin solid lines).
{\it (Right)} Constraints on the parameter $w$ of the cosmological
equation of state (thick solid lines from the combination of the 
method $A$ and $B$; dashed lines from the second method only; the crosses show 
the best-fit results, that is located at $\Omega_{\rm m}=0.35$ and $w=-1$ 
for the combined probability distribution).
The thin solid lines indicate
the constraints from SN Ia (P. Garnavich, priv. comm.; updated version 
of Garnavich et al. 1998 combining Riess et al. 1998 and Perlmutter
et al. 1999 dat sets).
The contours enclose the regions with $\Delta \chi^2=$2.30, 6.17, 11.8,
corresponding to 1, 2, 3 $\sigma$, respectively, for a distribution
with two degrees-of-freedom.
} \label{fgas:results} \end{figure*}

\section{CONCLUSIONS}

We show how the combined likelihood analysis of (i) the
representative value of $f_{\rm gas}$ in clusters of galaxies and (ii)
the requirement that $f_{\rm gas}(z)=$constant for
an assumed homogeneous class of objects with $T>4$ keV
can set stringent limits on the dark matter density
and any further contribution to the cosmic energy,
i.e. $\Omega_{\rm m}$ and $\Omega_{\Lambda}$ respectively.

First, a total $\chi^2$ distribution is obtained by combining
the two distributions presented in eqn.~\ref{chi2a} and ~\ref{chi2b},
i.e. $\chi^2 = \chi^2_A+\chi^2_B$. 
The resulting likelihood contours (Figure~\ref{fgas:results})
are obtained marginalizing over the range of parameters not investigated.
With further {\it a priori} assumptions on $\Omega_{\rm b}$ and $H_0$,
and assuming a flat geometry of the universe, we constrain
(see right panel in Figure~\ref{fgas:results})
the dark energy pressure-to-density ratio to be 
\begin{equation}
w < -0.82 (1 \sigma), -0.49 (2 \sigma), -0.17 (3 \sigma).
\end{equation}
This constraint is in excellent agreement with the bound on $w$ obtained
with independent cosmological datasets, such as the angular power spectra
of the Cosmic Microwave Background (e.g. Baccigalupi et al. 2002), 
the magnitude-redshift relation
probed by distant type Ia Supernovae and the power spectrum obtained
from the galaxy distribution in the two-degrees-field (for a combined
analysis of these datasets, see, e.g., Hannestad \& M\"ortsell 2002 and
reference therein). 
Moreover, this upper bound is completely in agreement with $w=-1$ 
as required for the equation of state of the ``cosmological constant''. 
Fixing $w=-1$, we obtain (for one interesting parameter)
\begin{equation} \begin{array}{l}
\Omega_{\rm m} = 0.34^{+0.03}_{-0.03} (1 \sigma), \ ^{+0.11}_{-0.05} (2 \sigma),
 \ ^{+0.17}_{-0.08} (3 \sigma)  \\
\Omega_{\Lambda} = 1.30^{+0.30}_{-0.46} (1 \sigma), \ ^{+0.44}_{-1.09} (2 \sigma),
 \ ^{+0.52}_{...} (3 \sigma).
\end{array}
\end{equation}

Finally, imposing a flat Universe (i.e. $\Omega_k =0$) as the recent constraints
from the angular power spectrum of the cosmic microwave background 
indicate (e.g. de Bernardis et al. 2002 and references therein), we obtain that
$\Omega_{\rm m} = 1-\Omega_{\Lambda} = 0.33^{+0.07}_{-0.05}$ at 95.4 per cent
confidence level.

Our limits on cosmological parameters fit nicely 
in the cosmic concordance scenario
(Bahcall et al. 1999, Wang et al. 2000), 
with a remarkable good agreement with independent
estimates derived from the angular power spectrum of 
Cosmic Microwave Background (Netterfield et al. 2002, Sievers et al. 2002), 
the magnitude--redshift relation
for distant supernovae type Ia (Riess et al. 1998, Perlmutter et al. 1999;
the likelihood region from a sample of SN type Ia as described in Leibundgut
2001 is shown in Figure~\ref{fgas:results}, panel at the bottom-left),
the power spectrum from the galaxy distribution in the 
2dF Galaxy Redshift Survey (e.g., Efstathiou et al. 2002)
and from galaxy clusters (e.g. Schuecker et al. 2002), 
the evolution of the X-ray properties of clusters of galaxies
(e.g. Borgani et al. 2001, Arnaud, Aghanim \& Neumann 2002, 
Henry 2002, Rosati, Borgani \& Norman 2002).
For example, combining the constraints in Figure~\ref{fgas:results} between
the allowed regions from the gas mass fraction and the 
magnitude--redshift relation for SN-Ia, we obtain 
(2 $\sigma$ statistical error) $\Omega_{\rm m}=0.34^{+0.07}_{-0.05}$ and
$\Omega_{\Lambda}=0.94^{+0.28}_{-0.32}$. These values are 
0.5 and 0.6 $\sigma$ higher, respectively, than the CMB constraints obtained
with the SN-Ia prior (see Table~4 in Netterfield et al. 2002).
Moreover, by combining $f_{\rm gas}$ and SN-Ia measurements
we can obtain a very tight constraint on $w$
(right panel in Figure~\ref{fgas:results}): $w < -0.89$ and
$\Omega_{\rm m}=0.32^{+0.05}_{-0.05}$ at the 95.4 per cent confidence level.

\begin{figure}
\epsfig{figure=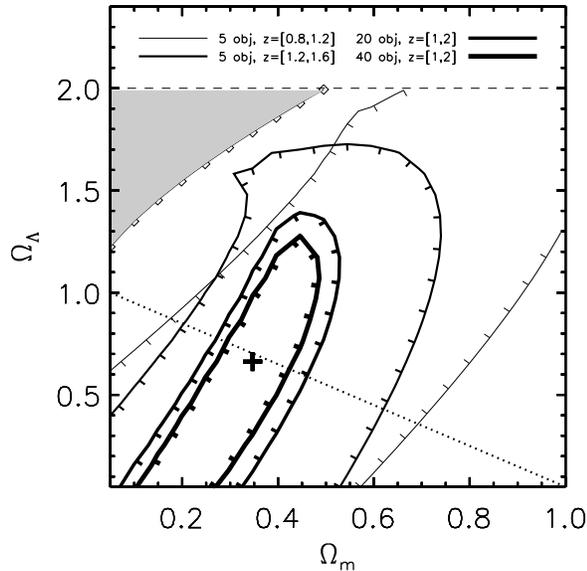,width=0.5\textwidth}
\caption{
Constraints in the $\Omega_{\rm m}-\Omega_{\Lambda}$ plane
as in Fig.~\ref{fig:dang} for simulated sets of gas mass fraction
measurements. The contours are at 1 $\sigma$ confidence level.
A cosmological model with $\Omega_{\rm m}=1-\Omega_{\Lambda}=0.34$ is assumed 
and only the $d_{\rm ang}^{3/2}$ dependence is considered.
In the redshift range $[1, 2]$, the increase by a factor of two of the sample
allows to reduce the upper limit on $\Omega_{\Lambda}$ by about 10 per cent.
} \label{fgas:simula} \end{figure}

We have demonstrated how the measurements of the cluster gas mass fraction 
represent a powerful tool to constrain the cosmological parameters and,
in particular, the cosmic matter density, $\Omega_{\rm m}$.
Nonetheless, the limits on $\Omega_{\Lambda}$ and $w$, though weaker, provide 
a complementary and independent estimate with respect to the most recent
experiments in this field.
On this item, it is worth noticing that our constraints on $\Omega_{\Lambda}$
are mostly due to the $d_{\rm ang}$ dependence of $f_{\rm gas}$
(cf. Fig.~\ref{fig:cosmo_cfr}). Thus, a larger sample of high$-z$ clusters
with accurate measurements of the gas mass fraction will significantly
shrink the confidence contours, as we show in Figure~\ref{fgas:simula}.
Compilations of such datasets will be possible in the near future 
using moderate-to-large area surveys obtained from observations 
with \chandra and \xmm satellites. 

\section*{ACKNOWLEDGEMENTS} 
We thank Stefano Borgani and Peter Schuecker for very useful discussions.
Bruno Leibundgut and Peter Garnavich are thanked for proving us 
the likelihood distributions 
from supernovae type Ia plotted in Figure~\ref{fgas:results}.
Pat Henry gave us a list of formula in advance of publication
that was useful to check our definition of the density contrast.
The anonymous referee is thanked for useful comments.
PT acknowledges support under the ESO visitor program in Garching.

\appendix
  \renewcommand{\thefigure}{A-\arabic{figure}}
  % redefine the command that creates the equation no.
  \setcounter{figure}{0}  % reset counter 

\begin{figure*}
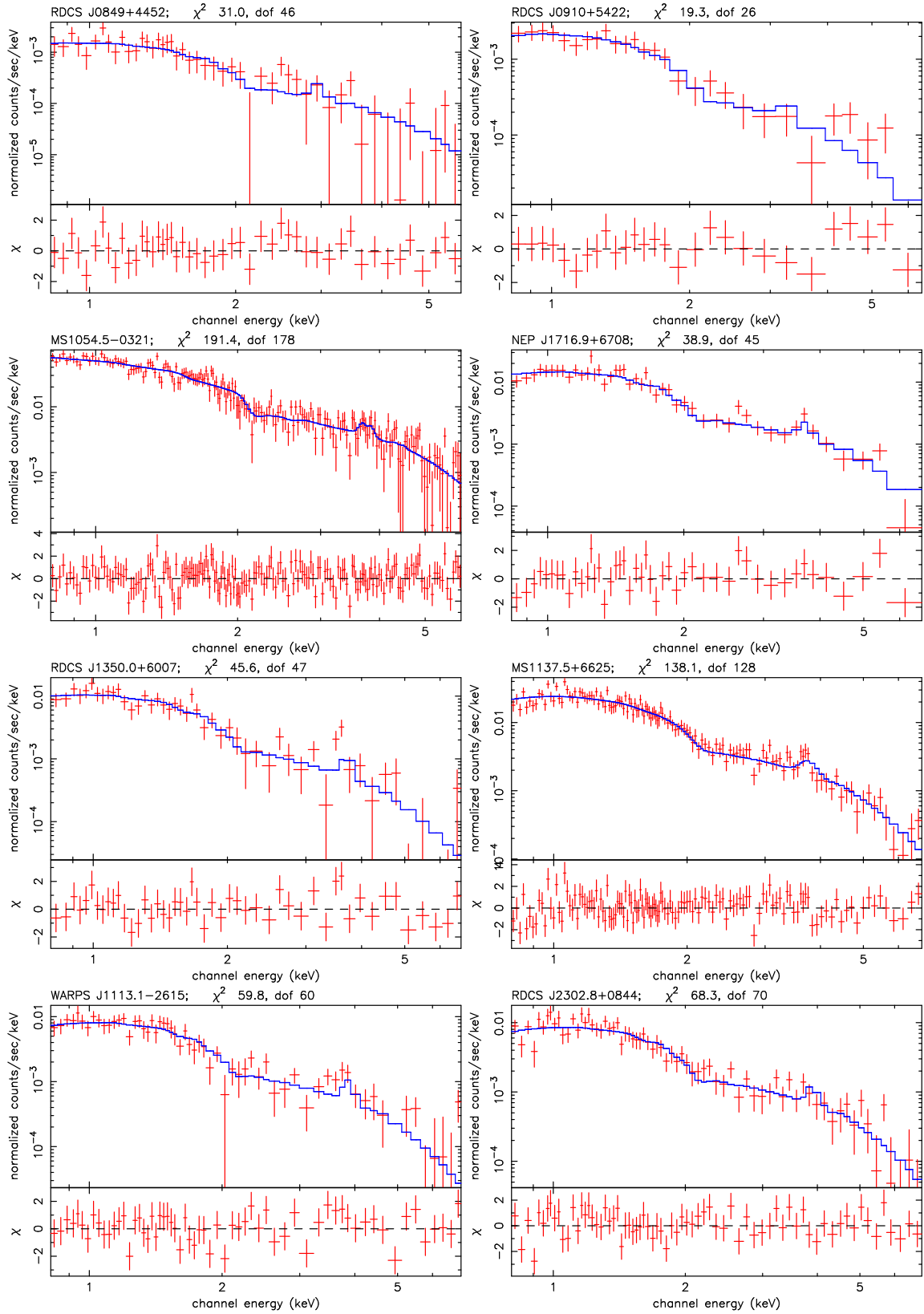

  \hbox{
  \epsfig{figure=ms3041f10a.ps,width=0.3\textwidth,angle=-90}
  \epsfig{figure=ms3041f10b.ps,width=0.3\textwidth,angle=-90}
  } \hbox{
  \epsfig{figure=ms3041f10c.ps,width=0.3\textwidth,angle=-90}
  \epsfig{figure=ms3041f10d.ps,width=0.3\textwidth,angle=-90}
  } \hbox{
  \epsfig{figure=ms3041f10e.ps,width=0.3\textwidth,angle=-90}
  \epsfig{figure=ms3041f10f.ps,width=0.3\textwidth,angle=-90}
  } \hbox{
  \epsfig{figure=ms3041f10g.ps,width=0.3\textwidth,angle=-90}
  \epsfig{figure=ms3041f10h.ps,width=0.3\textwidth,angle=-90}
  } 
\caption{Data and best-fit {\sc MEKAL} model of the spectrum 
of the galaxy clusters at high redshift in our sample.
} \label{data_model} 
\end{figure*}

\begin{figure*}
  \hbox{
  \epsfig{figure=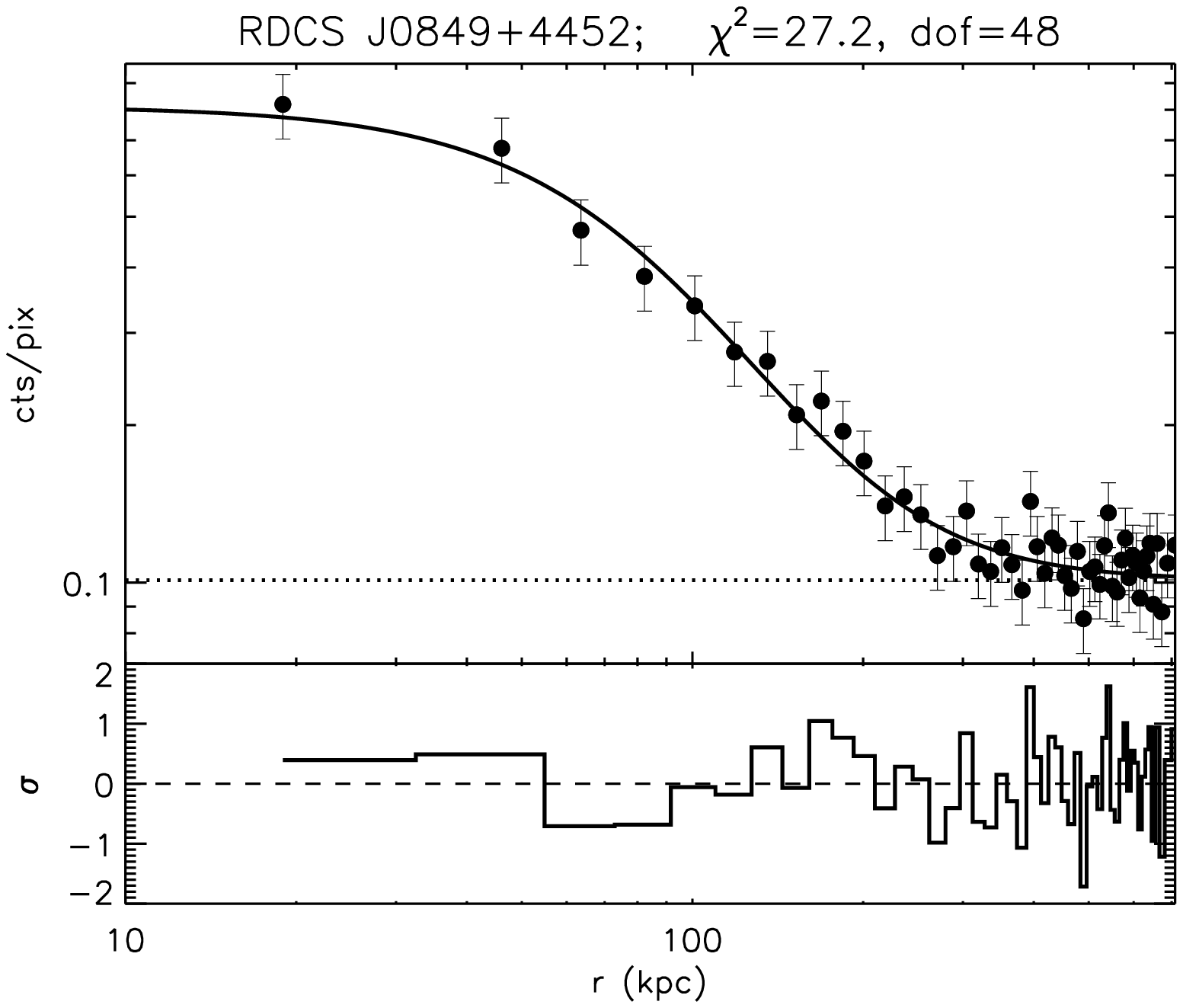,width=0.4\textwidth}
  \epsfig{figure=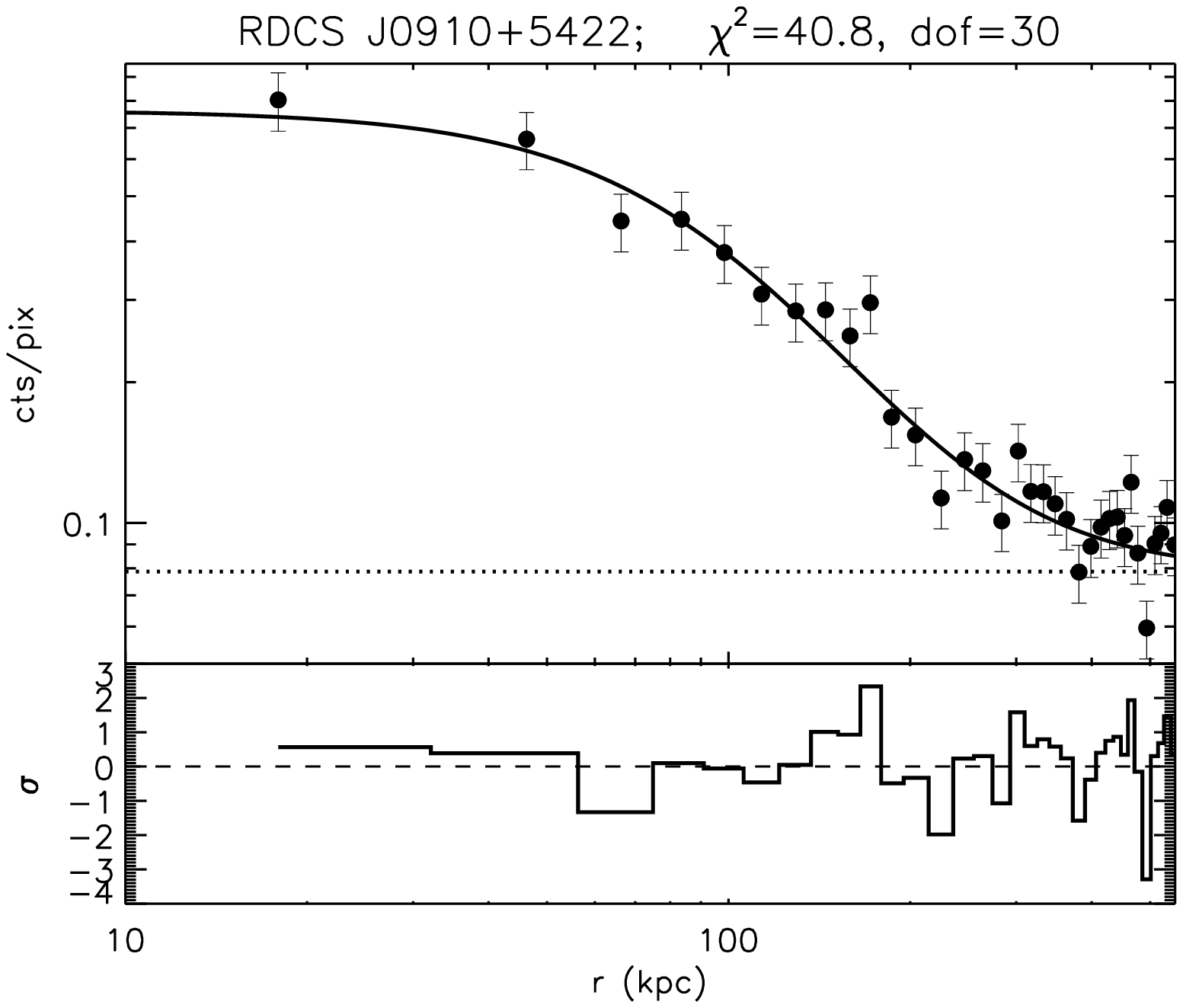,width=0.4\textwidth}
  } \vspace*{-0.5cm} \hbox{
  \epsfig{figure=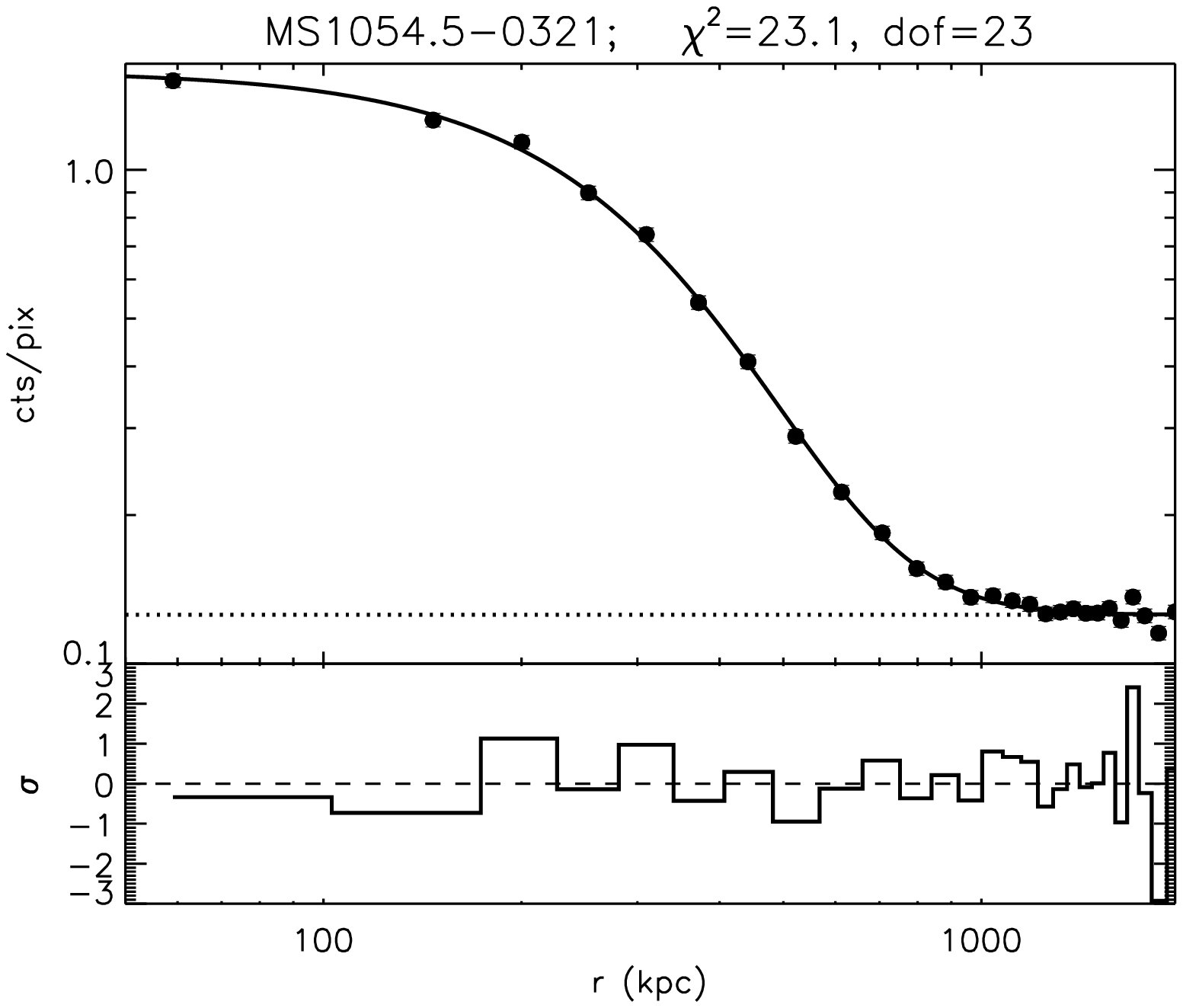,width=0.4\textwidth}
  \epsfig{figure=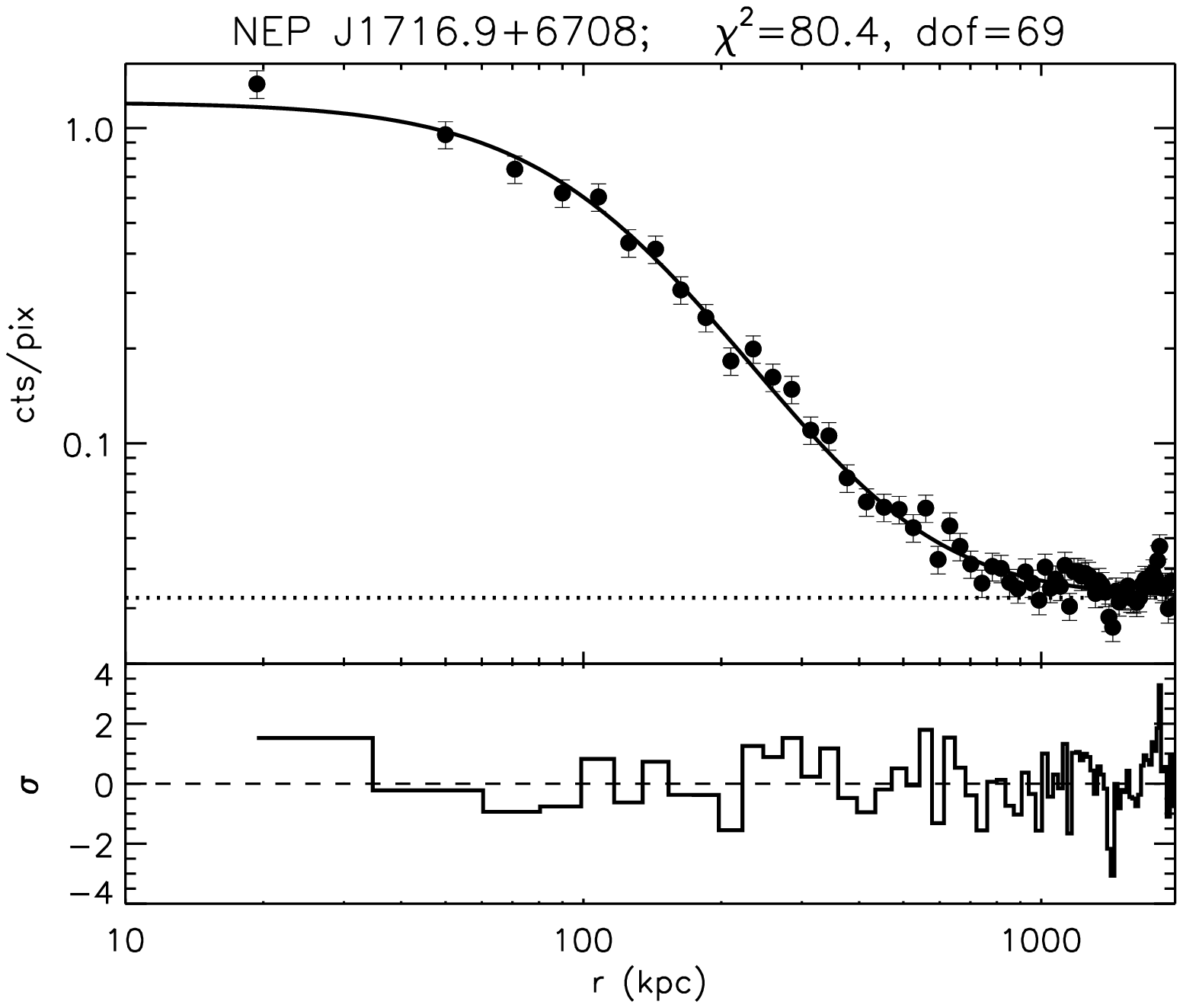,width=0.4\textwidth}
  } \hbox{
  \epsfig{figure=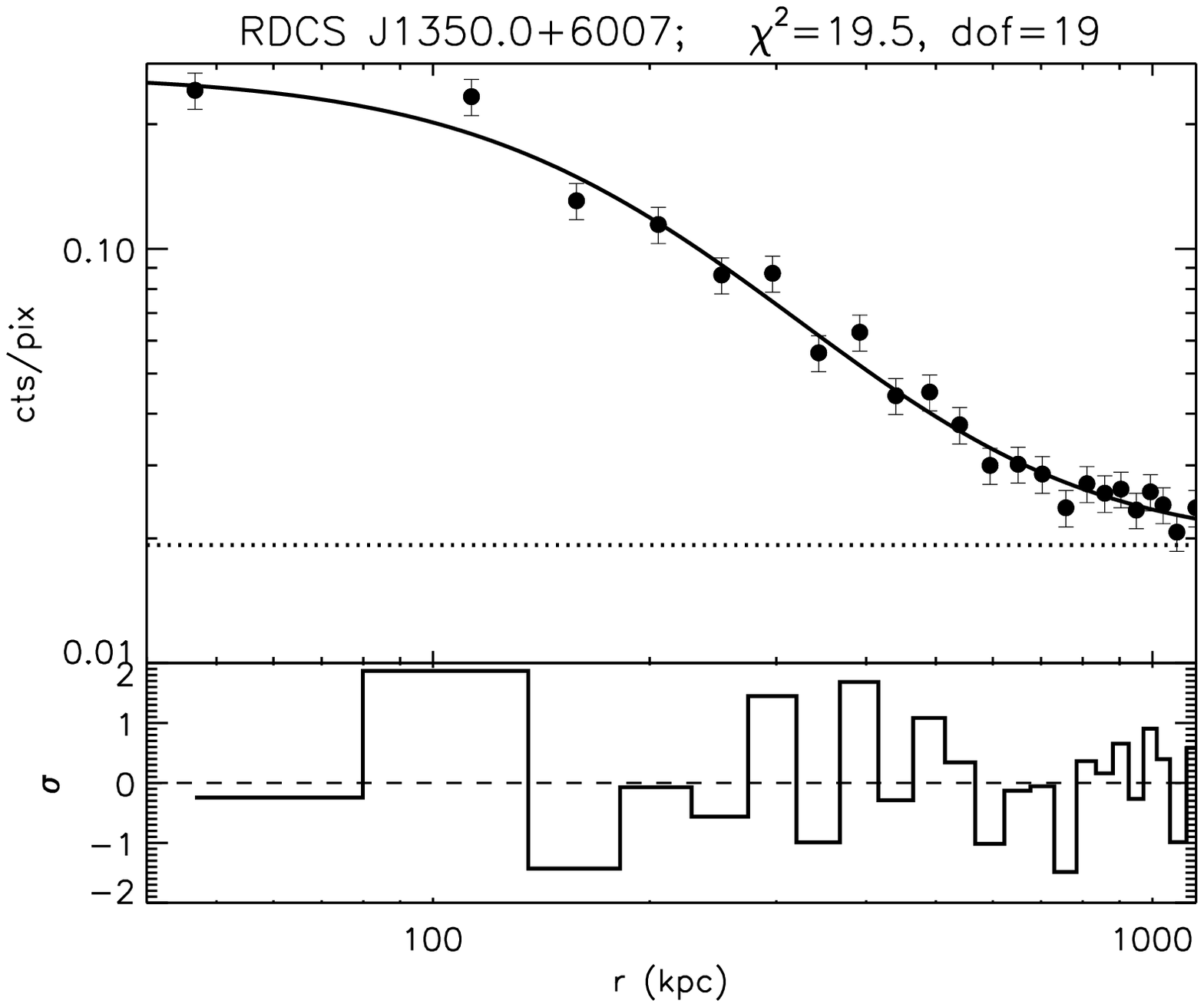,width=0.4\textwidth}
  \epsfig{figure=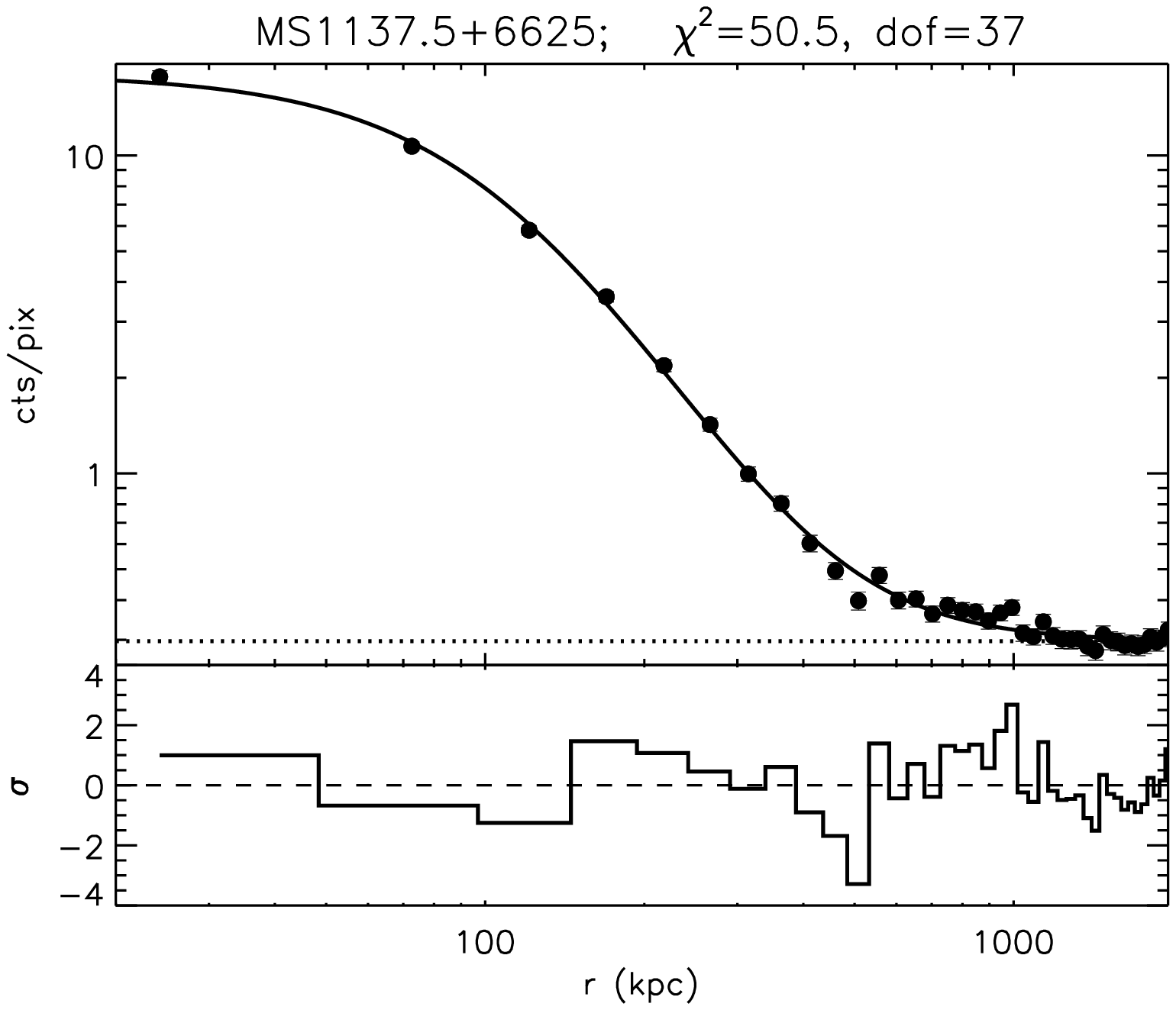,width=0.4\textwidth}
  } \hbox{
  \epsfig{figure=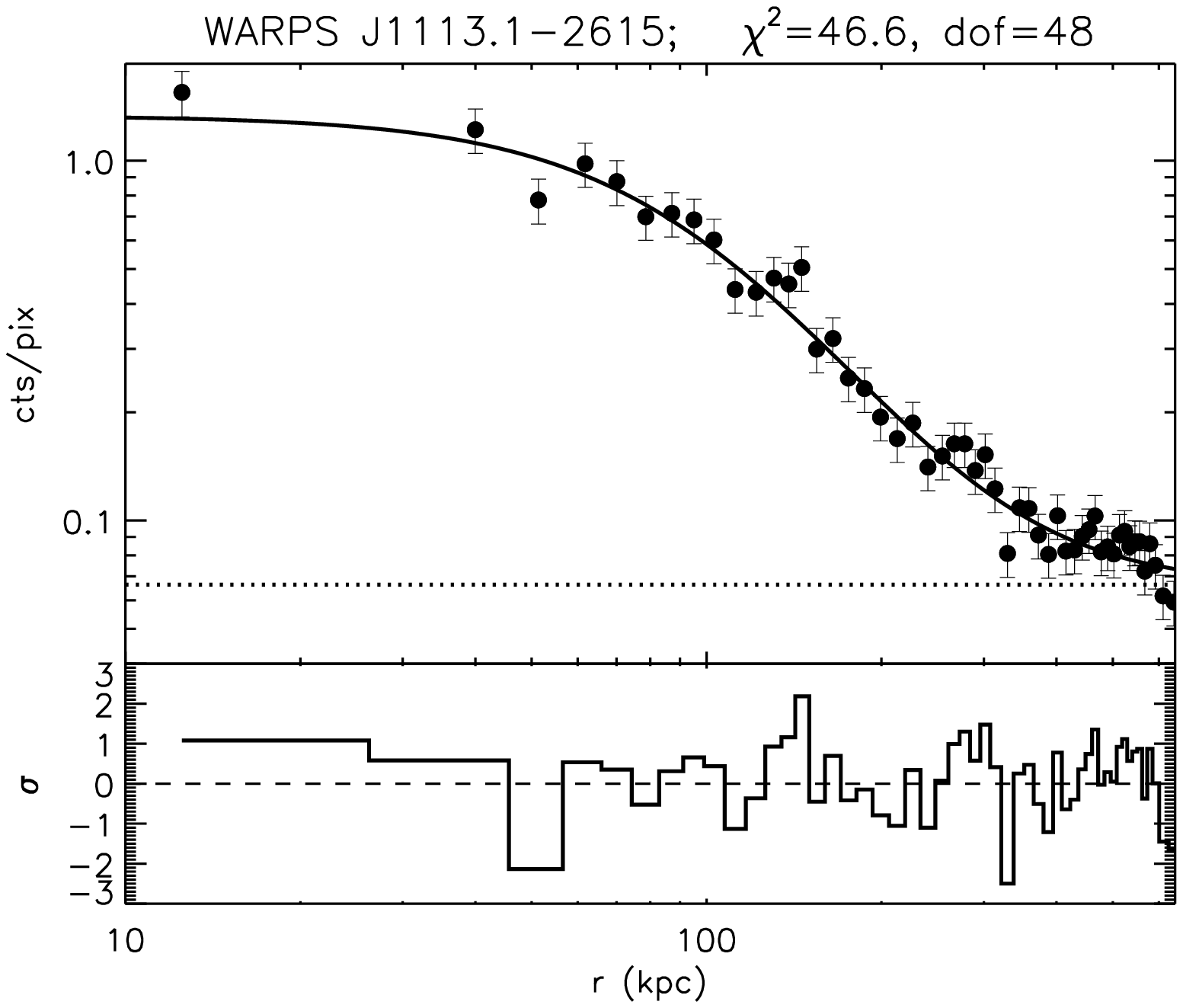,width=0.4\textwidth}
  \epsfig{figure=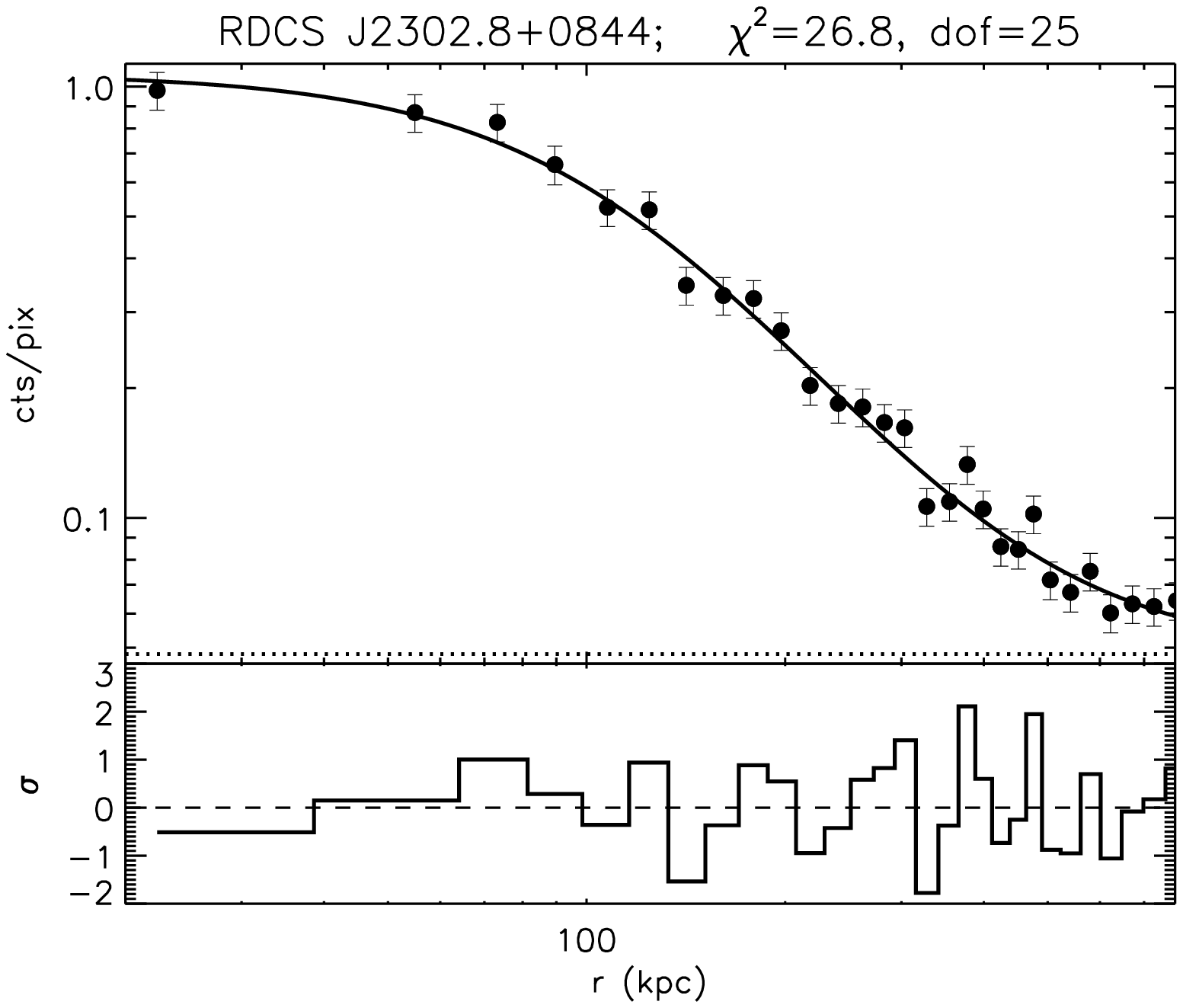,width=0.4\textwidth}
  } 
\caption{Data and best-fit $\beta$-model of the surface brightness profile of the 
galaxy clusters at high redshift in our sample. Dotted lines indicate the best-fit background
value.}  
\end{figure*}

\end{document}